\documentclass[11pt]{article}

\usepackage[latin1]{inputenc}

\usepackage{amstext,amsmath,amssymb,amsfonts,bbm,amsmath}
\usepackage{extarrows}
\usepackage{hyperref}
\usepackage{authblk}
\usepackage{caption} 
\usepackage{epsfig} 
\usepackage{color} 
\usepackage{graphicx} 
\usepackage{braket} 
\usepackage{slashed} 
\usepackage{dsfont} 
\usepackage{amsthm}  

\usepackage{cite}

\allowdisplaybreaks[4]

\usepackage{float}  
\usepackage{subfig}
\usepackage[lmargin=50pt,rmargin=60pt,tmargin=60pt,bmargin=65pt]{geometry}

\newcommand{\be}{\begin{equation}}
\newcommand{\ee}{\end{equation}} 
\newcommand{\Tr}{{\rm Tr}}
\newcommand{\tr}{{\rm tr}} 

\DeclareMathOperator{\im}{\mathrm{i}}

\newcommand{\mba}{\mathbf{a}}
\newcommand{\mbb}{\mathbf{b}}
\newcommand{\mbc}{\mathbf{c}}
\newcommand{\mbd}{\mathbf{d}}

\newcommand{\cG}{\mathcal{G}}

\newcommand{\cZ}{\mathcal{Z}}

\allowdisplaybreaks[4]

\theoremstyle{remark}

\begin{document}

\title{\bf Hints of unitarity at large $N$\\ in the $O(N)^3$ tensor field theory}

\author[1]{Dario Benedetti}
\author[1,2]{Razvan Gurau}
\author[1]{Sabine Harribey}
\author[1]{Kenta Suzuki}


\affil[1]{\normalsize \it 
 CPHT, CNRS, Ecole Polytechnique, Institut Polytechnique de Paris, Route de Saclay, \authorcr 91128 PALAISEAU, 
 France
 \authorcr
emails: dario.benedetti@polytechnique.edu, rgurau@cpht.polytechnique.fr, sabine.harribey@polytechnique.edu,
 kenta.suzuki@polytechnique.edu
 \authorcr \hfill }

\affil[2]{\normalsize\it 
Perimeter Institute for Theoretical Physics, 31 Caroline St. N, N2L 2Y5, Waterloo, ON,
Canada
 \authorcr \hfill}


\date{}

\maketitle

\hrule\bigskip

\begin{abstract}

We compute the OPE coefficients of the bosonic tensor model of \cite{Benedetti:2019eyl} for three point functions with two  fields and a bilinear with zero and non-zero spin.
We find that all the OPE coefficients are real in the case of an imaginary tetrahedral coupling constant, while one of them is not real in the case of a real coupling.
We also discuss the operator spectrum of the free theory based on the  character decomposition of the partition function.

\end{abstract}

\hrule\bigskip

\tableofcontents

\section{Introduction}
\label{sec:introduction}

Recently there has been extensive interest in tensor models because they admit a new kind of large $N$ limit, the \emph{melonic} limit \cite{critical,RTM,Klebanov:2018fzb,Prakash:2019zia}. The melonic limit is different from both the large $N$ limit of vector models \cite{Guida:1998bx,Moshe:2003xn} (dominated by bubble diagrams) and the one of matrix models\cite{'tHooft:1973jz,Brezin:1977sv,DiFrancesco:1993nw} (dominated by planar diagrams). Although as algebraic objects tensors are more complicated than matrices, their large $N$ limit is simpler because the melonic graphs are a subset of the planar graphs. The melonic limit is also obtained as a large $D$ limit of planar diagrams, or at large $N$ in matrix--tensor models\cite{Ferrari:2017ryl,Ferrari:2017jgw,Azeyanagi:2017mre}.

Tensor models were initially studied in zero dimension in the context of quantum gravity and random geometry \cite{Ambjorn:1990ge,Sasakura:1990fs,color,RTM,review}. They were then studied in one dimension \cite{Witten:2016iux,Gurau:2016lzk,Klebanov:2016xxf,Peng:2016mxj,Krishnan:2016bvg,Krishnan:2017lra,Bulycheva:2017ilt,Choudhury:2017tax,Halmagyi:2017leq,Klebanov:2018nfp,Carrozza:2018psc,Klebanov:2019jup,Ferrari:2019ogc} (see also \cite{Delporte:2018iyf,Klebanov:2018fzb} for reviews) as a generalization of the Sachdev-Ye-Kitaev model \cite{Sachdev:1992fk,Kitaev2015,Maldacena:2016hyu,Polchinski:2016xgd,Jevicki:2016bwu,Gross:2016kjj} without quenched disorder.

Tensor models can also be generalized in $d$ dimensions to proper field theories. In this setting they give rise to a new family of conformal field theories (CFTs) at large $N$ which are analytically accessible \cite{Giombi:2017dtl,Prakash:2017hwq,Benedetti:2017fmp,Giombi:2018qgp,Benedetti:2018ghn,Popov:2019nja}.
We call these conformal field theories \textit{melonic}.

In \cite{Benedetti:2019eyl}, a bosonic tensor field theory model has been shown to have an infrared attractive fixed point of the renormalization group flow.  Besides the tensorial structure of the interactions, already present for example in \cite{Giombi:2017dtl}, the model has two distinctive features: first, the kinetic term is non-local, or in other words, in the free limit we have a generalized free field theory; second, the coupling of the so-called \emph{tetrahedral} interaction is imaginary.
Although these two aspects of the model might seem somewhat exotic, they are not at all unprecedented.
Generalized free fields are a rather old idea \cite{Greenberg:1961mr}, and their interacting counterpart, in the case of a single scalar field, is known as the long-range Ising model \cite{Fisher:1972zz,Sak:1973}. The latter has been studied extensively with various methods, including constructive methods  \cite{Brydges:2002wq,Abdesselam:2006qg}, large-$N$ expansion \cite{Brezin:2014}, functional renormalization group \cite{Defenu:2014}, and CFT methods \cite{Paulos:2015jfa,Behan:2017emf}.
Our choice of kinetic term would correspond in the long-range Ising model to the transition point between the mean field theory behavior and the long-range one. The transition being continuous, our choice would then give mean field behavior in such model, i.e.\ no non-trivial fixed point. The existence of a non-trivial IR fixed point in our case is due mainly to the tetrahedral interaction.
Concerning its imaginary coupling, a famous example of field theory with imaginary coupling is given by the Lee-Yang model with an $\im \lambda \phi^3$ interaction \cite{Fisher:1978pf,Cardy:1985yy}, which is a real but non-unitary conformal field theory \cite{gorbenko2018walking}.

In this paper, we aim to further analyze the melonic CFT describing the fixed point of \cite{Benedetti:2019eyl}.
Although we have not proven that the theory is indeed conformal invariant, we are encouraged to think so on the basis of the conformal invariance of the long-range Ising model \cite{Paulos:2015jfa}, and the self-consistency of our CFT-based results. We will thus assume conformal invariance of the fixed-point theory, and work within the framework of CFT.
The main point we want to elucidate here is whether such a CFT is unitary. This question is more subtle than it looks at first sight.

A CFT is determined by the list of dimensions of primary operators and OPE coefficients among them.
All the correlation functions can then be obtained using the operator product expansion.
In \cite{Benedetti:2019eyl}, the dimensions of the primary bilinear operators with spin $J=0$ were obtained.
Here we complete this list and obtain the dimensions $h$ of the primary bilinear operators with arbitrary spin $J$:
\be
O_{h,J} \sim 
[ (\partial^2)^{\dots } \partial_{(\mu_1} \dots \partial_{\mu_i}\phi] [\partial_{\mu_{i+1}}\dots \partial_{\mu_J)} (\partial^2)^{\dots}\phi] -  {\rm traces}  \;.
\ee
We then compute an infinite (sub)set of OPE coefficients, namely those of the three point functions of two fundamental fields and a bilinear primary:
\begin{equation}
 \langle \phi \phi \,  O_{h,J}  \rangle \, .
\end{equation}
In order to completely characterize the CFT we would furthermore need the conformal dimensions and OPE coefficients of all the other primary operators. We leave this study for future work and below we focus on the bilinear primaries only.

A necessary (but not sufficient) condition for a CFT to be unitary is for all the OPE coefficients to be real \cite{gorbenko2018walking}\footnote{An additional requirement is for all the dimensions h of primaries to satisfy the unitarity bound $h\geq \frac{d}{2}-1$}.
As the tetrahedral coupling is purely imaginary at the fixed point,  we expect some OPE coefficients to be complex in our model and the CFT to not be unitary. 

However, a tantalizing possibility exists: it is possible that complex OPE coefficients pertain only to three point functions that are subleading in $N$. That is, it is possible that 
even though the finite $N$ CFT is not unitary, the large $N$ CFT is. 
The results of this paper point in this direction: all the OPE coefficients we compute are real in the case of an imaginary tetrahedral coupling.

\paragraph{Outline of results}

Our results are the following. We consider the $O(N)^3$ tensor model of \cite{Benedetti:2019eyl} in $d< 4$ dimensions. In the large $N$ limit, but non perturbatively in the coupling constants, this model has four lines of fixed points parametrized by the tetrahedral coupling constant $\lambda$. For purely imaginary $\lambda$, one of the fixed points is infrared attractive. 

First, we compute the dimensions of bilinear operators $O_{h,J}$.
We find two types of solutions at small renormalized tetrahedral coupling $g$. The first type:
\begin{equation}
h_{\pm} = \frac{d}{2} \pm 2 \frac{  \Gamma(d/4)^2  }{ \Gamma(d/2) } \sqrt{   - 3g^2  } + {\cal O}(g^3) \, ,
\end{equation}
only exists for the scalar (spin $J=0$) case. It is complex (at all orders in $g$) for real tetrahedral coupling and real (at all orders in $g$) for purely imaginary tetrahedral coupling. 
The second type:
\begin{equation}
 h_{m,J} = \frac{d}{2} + J + 2m -  \frac{ \Gamma(d/4)^4 
     \Gamma(m + J)  \Gamma(m+1 - \frac{d}{2}) \sin\left(\frac{\pi d}{2} \right)  }{ \Gamma(\frac{d}{2} + J + m ) \Gamma(m+1) \; \pi } 6g^2  + 
     {\cal O}(g^4) \, ,
\end{equation}
with $m,J \in \mathbb{N},~ (m,J)\neq (0,0)$ exists for both scalar $J=0$ and spin $J>0$. It is (at all orders in $g$) real for both real and purely imaginary tetrahedral coupling. 
In the free limit $g=0$, we recover the classical dimensions $\frac{d}{2}+J+2m$. 

Next, we computed the OPE coefficients $c_{m,J}$ of the three point functions $  \langle \phi \phi \, O_{h_{m,J},J}\rangle$. 
For $(m,J)\neq (0,0)$ the OPE coefficients $c_{m,J}$ are real (at all orders in $g$): 
\begin{align}
\label{OPE-coef-generalmJ}
 c_{m,J}^2 & = 2 
  \frac{  \Gamma(J+\frac{d}{2})   \Gamma(J+m )
    \Gamma(\frac{d}{2} + J + 2m  -1)
    \Gamma( 1 + m - \frac{d}{2}  )\Gamma(\frac{d}{4} + J + m  )^2
 }{ \Gamma(J+1) \Gamma(m+1 )
  \Gamma( 1 + 2m - \frac{d}{2}  )\Gamma( \frac{d}{2} + J + m    ) 
  \Gamma(J + 2m  )\Gamma(\frac{d}{2} + 2J + 2m   -1)
  \Gamma( \frac{d}{4} -m  )^2
 }  \nonumber \\
 &  \quad+ {\cal O}(g^2) \;.
\end{align}
In particular the $c_{m,J}$ are real at order $g^0$ which points to a unitary free theory. 

At spin $J=0$, the OPE coefficient $c_{0,0}$ is complex (at all orders in $g$) for $g$ real and real (at all orders in $g$) for $g$ purely imaginary:
\begin{equation}
 c_{0,0}^2 =   2  \pm  \sqrt{ -3g^2 }   \frac{4\, \Gamma(d/4)^2   }{\Gamma(\frac{d}{2})} \bigg[  2\Psi(d/4) - \Psi(d/2) -  \Psi(1)   \bigg]  
+ {\cal O}(g^3) \;.
\label{OPE-coef-general00}
\end{equation}

The cases $d=2$ and $d=1$ are special so we treat them separately. While the $d=1$ case turns out to be very similar to the $d=3$ one, the $d=2$ case is not. In fact the $d\to2$ limit is singular, a phenomenon we detail below.

Our results hold at all orders in $g$. This is due to the fact that both the dimensions and the OPE coefficients consist in a real constant term plus either a series in $g^2$ with real coefficients or $\sqrt{-g^2}$ times a series in $g^2$ with real coefficients.

\paragraph{Plan of the paper}

In section \ref{sec:the model}, we introduce and review the model, its RG flow, fixed points and the operator product expansion.
In section \ref{sec:OPE}, we compute the dimensions of the bilinear primary operators and the corresponding OPE coefficients for $d\ne 1, 2$.
In section \ref{sec:d=3}, we detail the case $d=3$.
As the cases $d=2$ and $d=1$ are special, we study them respectively in section \ref{sec:d=2} and \ref{sec:d=1}.
In section \ref{sec:character}, we use representation theory to derive the spectrum of bilinear primary operators in the free theory for integer dimension $d=3,2,1$.
Our conclusions are given in section \ref{sec:conclusions}.
In appendix \ref{app:measure}, we give some technical details. In appendix \ref{app:free}, we further comment on the free theory.
Lastly, in appendix \ref{app:original syk}, we review the OPE coefficients of the original SYK model and those of the conformal SYK model of Gross and Rosenhaus \cite{Gross:2017vhb}.

\section{The model and the operator product expansion}
\label{sec:the model}
We study the tensor model of \cite{Benedetti:2019eyl}, that is  the $O(N)^3$ tensor model of Klebanov and Tarnopolsky \cite{Klebanov:2016xxf} and Carrozza and Tanasa \cite{Carrozza:2015adg} (CTKT model) with a modified covariance.

We consider a real tensor field of rank $3$, $\phi_{a_1a_2 a_3}(x)$,
transforming under $O(N)^3$ with indices distinguished by the position and we denote $\mba = (a^1,a^2,a^3)$.
The action of the model is \footnote{Repeated indices are summed;
we work in $d$ dimensional Euclidean space; we denote $x,y$ and so on positions, $\int_x \equiv \int d^dx$ and
$p,q$ and so on momenta and $\int_p \equiv \int \frac{d^dp}{(2\pi)^d}$.}:
\be\label{eq:action} 
\begin{split}
    S[\phi]  & =   \frac{1}{2} \int d^dx \;   \phi_{\mba}(x) (   - \partial^2)^{\zeta}\phi_{\mba}(x) + S^{\rm int}[\phi]\;,\crcr
   S^{\rm int}[\phi]  & = \frac{ m^{2\zeta}}{2} \int d^dx \;   \phi_{\mba}(x) \delta_{\mba \mbb} \phi_{\mbb}(x)  + 
   \frac{ \lambda }{4 N^{3/2}} \int d^d x \;   \delta^t_{\mba \mbb\mbc\mbd} \; \phi_{\mba}(x) \phi_{\mbb}(x)  \phi_{\mbc}(x) \phi_{\mbd }(x)\crcr
       & \qquad +   \int d^d x \;  \left( \frac{ \lambda_p }{4 N^{2}} \; \delta^p_{\mba\mbb; \mbc\mbd} +  \frac{ \lambda_d }{4 N^{3}}  \; \delta^d_{\mba\mbb; \mbc\mbd } \right) \; \phi_{\mba}(x) \phi_{\mbb}(x)  \phi_{\mbc}(x) \phi_{\mbd }(x) 
     \; ,     
\end{split}
\ee 
where $\partial^2=  \partial_{\mu}\partial^{\mu}$,  $\delta_{\mba \mbb}  = \prod_{i=1}^3 \delta_{a^i b^i} $ and:
\begin{align}
    \delta^t_{\mba \mbb\mbc\mbd}  = \delta_{a^1 b^1}  \delta_{c^1 d^1} \delta_{a^2 c^2}  \delta_{b^2 d^2 } \delta_{a^3 d^3}   \delta_{b^3 c^3} \;  , \quad
  \delta^p_{\mba\mbb; \mbc\mbd }= \frac{1}{3} \sum_{i=1}^3  \delta_{a^ic^i} \delta_{b^id^i} \prod_{j\neq i}  \delta_{a^jb^j}  \delta_{c^jd^j} \;,
 \quad  \delta^d_{\mba\mbb; \mbc\mbd }  = \delta_{\mba \mbb}  \delta_{\mbc \mbd} \;, 
\end{align}
where $t$ stands for \emph{tetrahedron}, $d$ for \emph{double-trace} and $p$ for \emph{pillow} pattern of contraction of indices. 
We have not assign any subscript to the coupling $\lambda$ of the tetrahedral
invariant.

\begin{figure}[ht]
\begin{center}
\includegraphics[width=0.5\textwidth]{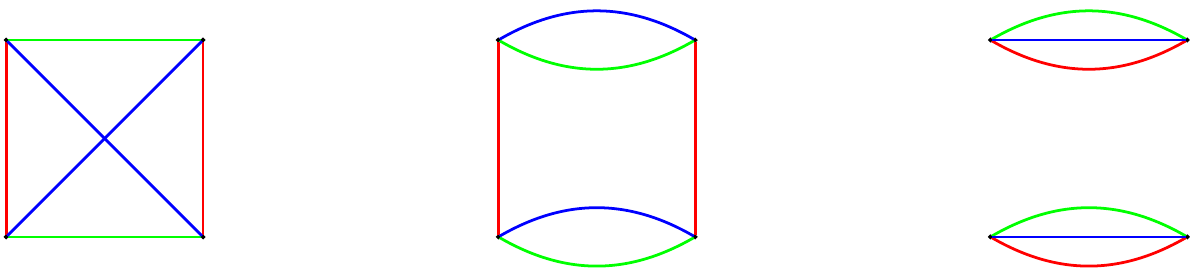}
 \caption{Graphical representation of the quartic $O(N)^3$ invariants. From left to right: the tetrahedron, the pillow, and the double-trace (there are three pillow contractions, distinguished by the color of the vertical edge).} \label{fig:interactions}
 \end{center}
\end{figure}

As usual, it is convenient to introduce a graphical representation of the $O(N)^3$ invariants, which also justifies their names. We represent every tensor ($\phi_{\mba}$, $\phi_{\mbb}$ and so on) as a three valent vertex and every contraction of two indices ($a^i$ and $b^i$ for instance) as an edge with a color $i=1, 2$ or 3 (red, green and blue) corresponding to the position $i$ of the indices. The quartic invariants of Eq.~\eqref{eq:action} are represented in Fig.~\ref{fig:interactions}.

The difference between this model and the CTKT model is that the Laplacian is allowed to have a non integer power $\zeta \leq 1$. 
This modification preserves the reflection positivity of the propagator: the free theory is unitary for any $\zeta \le 1$. We set $\zeta = \frac{d}{4} $, rendering the quartic invariants marginal in any $d$  \cite{Benedetti:2019eyl}.

\begin{figure}[ht]
\begin{center}
\includegraphics[width=0.3\textwidth]{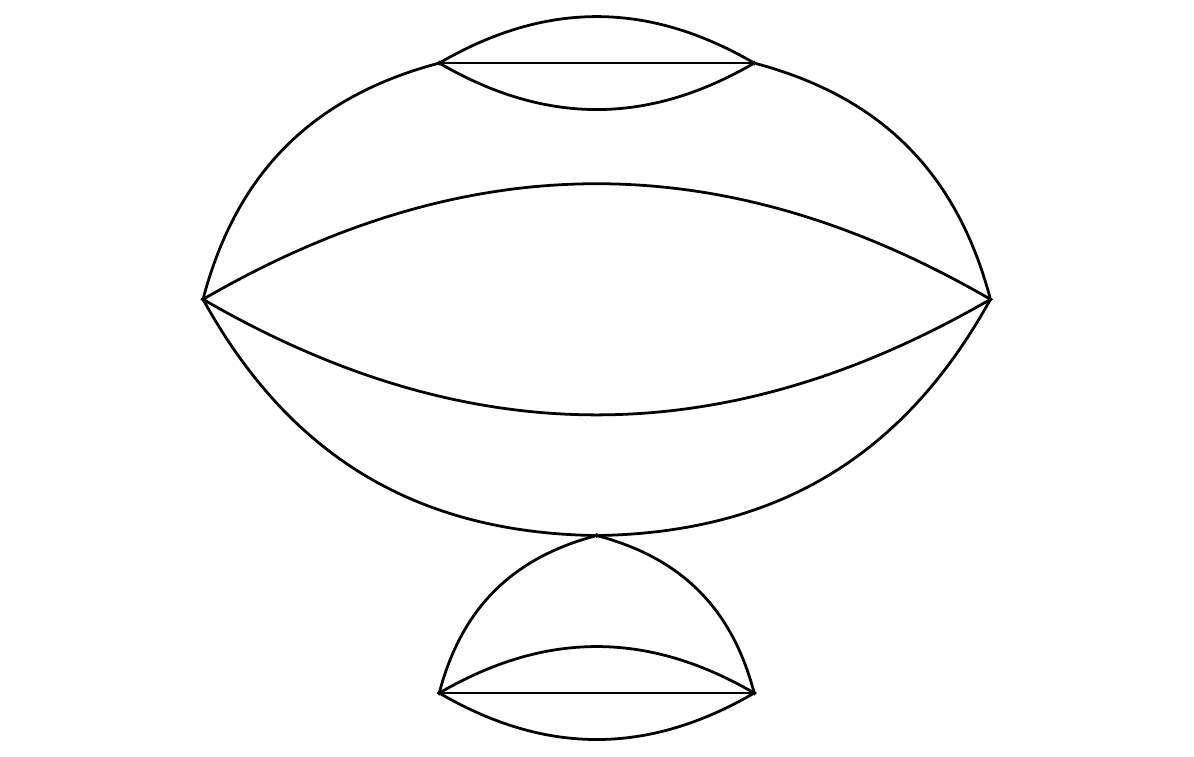}
 \caption{A melon tadpole graph, where all the invariants have been shrunk to point-like vertices.} \label{fig:melontadpoles}
 \end{center}
\end{figure}

We can expand the free energy and the connected $n$-point functions into connected Feynman graphs. Each interaction invariant is a $3$-colored graph and the Feynman propagators are represented as edges with a new color which we call 0 (pictured in black) connecting the tensors. This leads to a representation of the Feynman graphs as 4-colored graphs. It is sometimes convenient to simplify the graphs by shrinking each interaction invariant to a point. 
This model has a $1/N$ expansion\cite{Carrozza:2015adg,Klebanov:2016xxf} dominated by \emph{melon tadpole} graphs  \cite{Benedetti:2019eyl} with melons based on couples of tetrahedral vertices and tadpoles based on either pillow or double-trace vertices (see Fig.\ref{fig:melontadpoles}). 

The two point function is diagonal in the tensor indices: denoting $A=(\mba,x)$ and so on, the effective two-point function is: 
\be
G_{AB}=\braket{\phi_{\mba}(x) \phi_{\mbb}(y)} 
=\delta_{\mba \mbb} \, G(x-y) \;.
\ee

\paragraph{Renormalization.} 
For $\zeta=d/4$ the bare covariance reproduces the infrared scaling of the two-point function.  With this choice, in the large $N$ limit but non perturbatively in the coupling constants, the RG flow of the model has four lines of fixed points parameterized by the tetrahedral coupling $\lambda$. In detail, we get the following.

\emph{Two point function.} For any bare couplings there exists a choice of the bare mass such that the renormalized mass is tuned to zero. The wave function renormalization $Z$ is a finite rescaling and the Fourier transform of $G(x-y)$ is:
\be
  G(p)  = 
  \frac{1}{Z p^{2\zeta}} \, , \qquad   Z^4 - Z^3  =   \lambda^2 \frac{1}{(4\pi)^d } \;\frac{\Gamma \left( 1 -\frac{d}{4} \right) }{ \frac{d}{4}\Gamma\left( 3 \frac{d}{4}\right)} \, .
 \ee
 
\begin{figure}[ht]
\begin{center}
\includegraphics[width=0.5\textwidth]{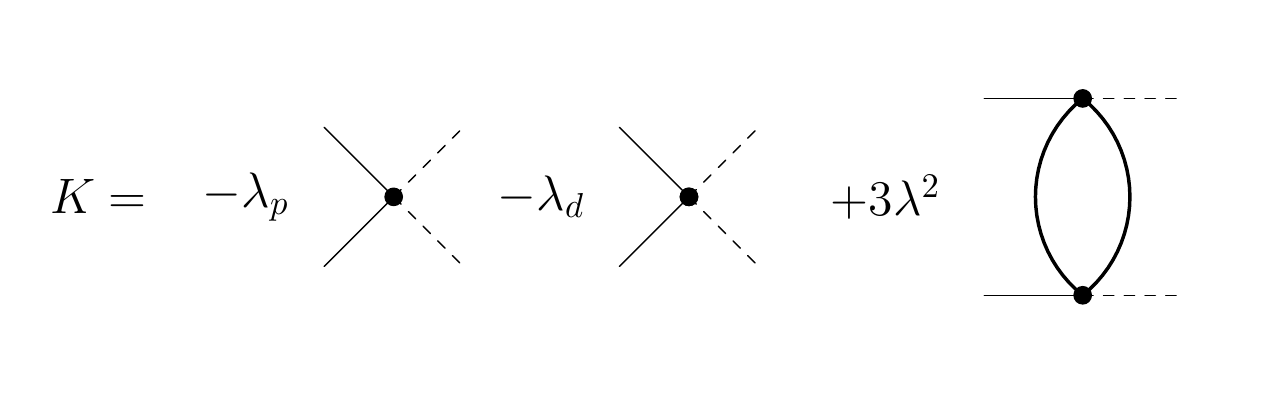}
 \caption{Graphical representation of the kernel amputated to the right at leading order in $N$. The first two terms are based respectively on pillow and double-trace vertices while the last one is based on a pair of tetrahedral vertices.} \label{fig:kernel}
 \end{center}
\end{figure}

 \emph{Four point function.}
The four point function is computed
\cite{Benedetti:2019eyl} in terms of the two particle irreducible four point kernel $\mathcal{K}$ and the projector on symmetric functions
$ {\cal S} $ \cite{Benedetti:2018goh}:
\begin{equation}\label{eq:2.0}
 \braket{\phi_{A}  \phi_{B} \phi_{C} \phi_{D} }^c  \, = \, 2 \left( \frac{{\cal K}}{1 - {\cal K}} {\cal S} \right)_{ AB; C'D' }  G_{C'C}  G_{D'D}  \; ,
\end{equation}
and at leading order in $N$ the four point kernel is (see Fig.\ref{fig:kernel}):
\begin{align}
  {\cal K}_{ (\mba,x')(\mbb,y') ; (\mbc,z)(\mbd,w) } = & G_{x'x} G_{y'y}  \bigg[  
    -  \lambda_p   \delta_{xy} \delta_{xz} \delta_{xw} \hat \delta^p_{\mba\mbb ; \mbc\mbd} -  \lambda_d  \delta_{xy} \delta_{xz} \delta_{xw} \hat \delta^d_{\mba\mbb ; \mbc\mbd}   
    + 3 \lambda^2 G_{xy}^2 \delta_{xz}\delta_{yw} \hat \delta^p_{\mba\mbb ; \mbc\mbd} 
 \bigg] \; ,
\end{align} 
where repeated positions are integrated, $G_{xy}\equiv G(x-y)$ and $\hat \delta^{p}_{\mba \mbb ; \mbc \mbd} = \frac{1}{N^2} \delta^{p}_{\mba \mbb ; \mbc \mbd}$ respectively $\hat \delta^d_{\mba \mbb ; \mbc \mbd} = \frac{1}{N^3} \delta^{d}_{\mba \mbb ; \mbc \mbd}$ are the rescaled pillow and double-trace contraction operators. 

With $\lambda_1=\lambda_p/3$ and $\lambda_2=\lambda_p+\lambda_d$, the four-point kernel in momentum space writes:
\begin{align}\label{eq:newker}
   K_{ p_1,p_2; q_1,q_2 } = & (2\pi)^d \delta(p_1+p_2 - q_1 -q_2)   G(p_1) G(p_2)  \crcr
&\qquad \bigg[ \bigg( \lambda^2 \int_q G(q)G(q+p_1-q_1)       -\lambda_1 \bigg) P_1 + \bigg( 3\lambda^2 \int_q G(q)G(q+p_1-q_1) -  \lambda_2 \bigg) P_2 
\bigg]\; ,
\end{align}
with $P_1 = 3( \hat \delta^p - \hat \delta^d )$ and $P_2 = \hat \delta^d$. 

\emph{Fixed points.} The tetrahedral coupling has a finite flow: the renormalized tetrahedral coupling is just a rescaling of the bare one by the wave function constant:
\begin{equation}
g \, = \, Z^{-2}\lambda \, , \qquad  \beta_g=k\frac{\partial g}{\partial k} \, = \, 0 \, .
\end{equation}

The $\beta$ functions of the renormalized couplings $g_1$ and $g_2$ are independent and quadratic \cite{Benedetti:2019eyl}:
\be
 \beta_{g_i} = k\frac{\partial g_i }{\partial k} \, = \beta_2^{(i)} g_i^2 -
  2\beta_1^{(i)} g_i + \beta_0^{(i)} \;, \qquad i\text{ not summed,} 
\ee
where the coefficients $\beta_{\dots}^{(i)} $ are power series in $g^2$. Each $\beta$ function admits two fixed points parametrized by $g$ and we get four lines of fixed points. 
For a purely imaginary tetrahedral coupling $g$, the fixed points are real and one of them is infrared attractive. In the rest of the paper we are interested in studying the CFT defined at this infrared attractive fixed point. 

\paragraph{Operator product expansion.}
The four point function in a CFT with a real field of dimension $\Delta_{\phi}$ can always be written as \cite{Simmons-Duffin:2017nub,Liu:2018jhs,Gurau:1907}:
\begin{equation} \label{eq:4pt}
\begin{split}
 \langle{\phi(x_1) \phi(x_2) \phi(x_3) \phi(x_4)} \rangle
 = & \langle{\phi(x_1) \phi(x_2) \rangle \; \langle \phi(x_3) \phi(x_4)} \rangle + \crcr
 & \qquad  +  \sum_J 
  \int_{\frac{d}{2}-\imath \infty}^{\frac{d}{2}+\imath\infty} \frac{dh}{2\pi \imath}
  \;\frac{1}{1-k(h,J)} \; \mu_{\Delta_{\phi}}^d(h,J)
     G^{\Delta_{\phi}}_{h,J}(x_i)\;,
\end{split}
\end{equation}
with $ G^{\Delta_{\phi}}_{h,J}(x_i) $ the conformal block \cite{Simmons-Duffin:2017nub,Liu:2018jhs} (a ``known'' function), $\mu_{\Delta_{\phi}}^d(h,J)$ the measure:
\begin{align}\label{eq:measure}
\mu_{\Delta_{\phi}}^d(h, J)  \, = &  \,  b^2 \left( \frac{ 1 + (-1)^J }{2} \right)
   \frac{  \Gamma(J+\frac{d}{2}) 
  }
  {  \Gamma(J+1)} 
   \crcr
&  \ \times    \frac{
 \Gamma( \frac{d}{2} - \Delta_{\phi})^2 
    \Gamma(\frac{ 2\Delta_{\phi} -d +h+J}{2}) 
    \Gamma(\frac{2\Delta_{\phi}-h+J}{2})
    \Gamma(h-1)\Gamma(d-h+J)\Gamma(\frac{h +J}{2})^2
 }{ \Gamma( \Delta_{\phi})^2 
  \Gamma(\frac{2d-2\Delta_{\phi}-h+J}{2})\Gamma(\frac{d-2\Delta_{\phi} +h+J}{2}) 
  \Gamma(h-\frac{d}{2})\Gamma(h+J-1)\Gamma(\frac{d- h  +J}{2})^2
 }  \;,
\end{align}
and $k(h,J)$ the eigenvalues of the two particle irreducible four point kernel \cite{Gurau:1907}. 
$b$ is the coefficient of the two-point function in direct space, but in the following we will omit it from the measure, meaning that we actually consider the four-point function for the rescaled fields $\phi/\sqrt{b}$.
The case $d=1$ is special \cite{Maldacena:2016hyu}: in that case one gets extra contours for $h$ around the even integers.

The OPE expansion is obtained by deforming the integration contour to the right and picking up the poles in the integrand. All the poles coming from the measure and the conformal block are spurious \cite{Simmons-Duffin:2017nub} hence:
\begin{equation}
  \langle{\phi(x_1) \phi(x_2) \phi(x_3) \phi(x_4)} \rangle
 =  \langle{\phi(x_1) \phi(x_2) \rangle \; \langle \phi(x_3) \phi(x_4)} \rangle 
+  \sum_{m,J} c_{m,J}^2  \; G^{\Delta_{\phi}}_{h_m,J}(x_i) \;,
\end{equation}
where $h_{m,J}$ are the poles of $(1-k(h,J))^{-1}$ \cite{Gurau:1907} and the squares of the OPE coefficients are the residues at the poles:
\begin{equation}\label{eq:OPE coef}
c_{m;J}^2=-\mu_{\Delta_{\phi}}^d(h_{m,J} ,J) \text{Res}\left[\frac{1}{1-k(h,J)}\right]_{h=h_{m,J} } 
=\frac{\mu_{\Delta_{\phi}}^d(h_{m,J},J)}{ k'( h_{m,J},J) }  \;,
\end{equation}
where the prime denotes a derivative respect to $h$. As we deform the contour to the right, we only pick up the poles with ${\rm Re}(h_{m,J}) \ge d/2$. If some poles lie on the original integration contour, that is ${\rm Re}(h_{m,J}) = d/2$, extra care must be taken.

From now on we focus on $d<4$. While the interesting values are $d=1,2$ and most especially $d=3$, all our results apply also in non integer $d$.

\section{Primary operators and OPE coefficients}
\label{sec:OPE}
We now compute the dimensions of the bilinear primaries of arbitrary spin and their OPE coefficients.

\paragraph{Dimensions of primaries.}
Our model has $\Delta_{\phi}=d/4$  \cite{Gurau:1907} and the eigenvalues of the two particle irreducible four point kernel \cite{Gurau:1907} are:
\begin{equation}
k(h,J)=3g^2\Gamma(d/4)^4
 \frac{\Gamma(-\frac{d}{4}+\frac{h+J}{2})\Gamma(\frac{d}{4}-\frac{h-J}{2})}{\Gamma(\frac{3d}{4}-\frac{h-J}{2})\Gamma(\frac{d}{4}+\frac{h+J}{2})} \;.
\label{eq:k(h,J)}
\end{equation}

We are interested in the solutions $h_{m,J}$ of the equation $k(h,J) = 1 $ with ${\rm Re}(h_{m,J}) \ge d/2$ for small $g$. Such solutions correspond to values of $h$ for which the ratio of Gamma functions diverges. As the Gamma function does not have any zeros in the complex plane ($1/\Gamma(z)$ is an entire function) divergences only arise near the poles of 
the numerator $ \Gamma(-\frac{d}{4}+\frac{h+J}{2})\Gamma(\frac{d}{4}-\frac{h-J}{2}) $. We are only interested in the poles in the region ${\rm Re}(h) \ge d/2$, therefore the relevant poles of the numerator are located at the classical dimensions of the bilinear spin $J$ operators:
\begin{equation}
h^{\rm classical} = d/2 + J + 2m  \; , \qquad m\ge 0 \;.
\end{equation}
The poles of $ \Gamma(-\frac{d}{4}+\frac{h+J}{2})$ are located at $h = d/2-J-2m$ which does not obey ${\rm Re}(h) \ge d/2$ for $(m,J)\neq (0,0)$. In the neighborhood  of each pole we parametrize $h = d/2 + J + 2m + 2z$ with $z\sim o(g)$. Then $k(h,J)$ becomes:  
\begin{equation}
k(d/2 + J + 2m + 2z  ,J) = 3g^2 \Gamma(d/4)^4 
    \frac{ \Gamma(J+m+z) \Gamma(-m-z)  }
    { \Gamma( \frac{d}{2} -m-z ) \Gamma( \frac{d}{2} + J +m + z) } \; .
\end{equation}

For any $d$, the pole $(m,J) = (0,0) $ is special because both the $\Gamma$ functions in the numerator are singular, while for the poles $(m,J)\neq(0,0)$ only one of  them is. Moreover the case $d=2$ is special as $\Gamma( \frac{d}{2} -m-z ) $ diverges for $m\ge 1$ at small $z$. We take $d\neq 2$ and we will deal with $d=2$ in section \ref{sec:d=2}.
The function $k(h,J)$ close to the pole $(m,J)$ takes the form:
\begin{equation}
\label{eq:2ks}
\begin{split}
  k_{(0,0)}(z) \, &= \, 3g^2 \Gamma(d/4)^4  \; 
   \frac{ \Gamma(1+z) \Gamma(1-z)  }{  (- z^2) \Gamma(\frac{d}{2}-z) \Gamma(\frac{d}{2}+z) }  \;, \crcr
k_{(m,J)} (z) \, & \xlongequal{(m,J) \neq (0,0)} \,
  3g^2 \Gamma(d/4)^4
    \frac{\Gamma(J+m+z)}{ \Gamma(\frac{d}{2}+J+m+z) }
     \; \frac{ \Gamma(1+z) \Gamma(1-z) }{ z \Gamma(m+1 + z) }
     \;\frac{ \Gamma( m+1 - \frac{d}{2} +z )}{  
       \Gamma(z-\frac{d}{2}) \Gamma(\frac{d}{2} +1 -z)}  \;,  
\end{split}
\end{equation}
where in the second line we used $\Gamma(-m + a) = (-1)^{m+1} \Gamma(-a) \Gamma(1+a) /\Gamma(m+1-a)$. The dimensions of the physical operators in the interacting theory are the solutions of the equation $k(h_{m,J},J)=1$, that is:
\begin{equation}
   h_{m,J} = \frac{d}{2} + J +2 m + 2z_{m,J}\;, \qquad
     k_{(m,J)} ( z_{m,J} ) = 1 \;.
\end{equation}

The solutions $z_{m,J}$ (which are the anomalous scalings of the bilinear primaries at the fixed point) are obtained as follows.

\paragraph{\it The case $(m,J) = (0,0)$} The anomalous dimension $z_{0,0}$ is the solution of:
  \begin{equation}\label{eq:z00}
     (- z^2) \frac{ \Gamma(\frac{d}{2}-z) \Gamma(\frac{d}{2}+z) }  
   { \Gamma(1+z) \Gamma(1-z)  }
    =    3g^2 \Gamma(d/4)^4  \; ,
  \end{equation}
  where only the solutions with ${\rm Re}(z)\ge 0$ are picked up.
  Observe that the left hand side of eq.~\eqref{eq:z00} is a series 
  in $z^2$ with real coefficients, therefore $z_{0,0}^2$ is a series in $g^2$ with real coefficients starting at first order which implies:
  \begin{equation}
    z_{0,0} =  \pm \frac{  \Gamma(d/4)^2  }{ \Gamma(d/2) } \sqrt{   - 3g^2  } 
       \left( 1 + \sum_{q\ge 0} C_q g^{2q}\right) \;, \qquad 
         C_q \in \mathbb{R} \;.
  \end{equation}
  At first order in $g$ we have:
  \begin{equation}
     z_{0,0} = \pm \frac{  \Gamma(d/4)^2  }{ \Gamma(d/2) } \sqrt{   - 3g^2  } + {\cal O}(g^3) \;, \qquad       h_{\pm} = \frac{d}{2} \pm 2 \frac{  \Gamma(d/4)^2  }{ \Gamma(d/2) } \sqrt{   - 3g^2  } + {\cal O}(g^3) \;.
  \end{equation}  
  
\paragraph{\it The case $(m,J)\neq (0,0)$} The remaining anomalous dimensions are the solutions of:
   \begin{equation}
    z \;  \frac{ \Gamma(\frac{d}{2}+J+m+z)  \Gamma(m+1 + z) \Gamma(z-\frac{d}{2}) \Gamma(\frac{d}{2} +1 -z)} 
      {\Gamma(J+m+z) \Gamma(1+z) \Gamma(1-z)
      \Gamma( m+1 - \frac{d}{2} +z )
      }
      =   3g^2 \Gamma(d/4)^4 \;.
   \end{equation}
   Obviously $z_{m,J}$ are series in $g^2$ with real coefficients and at first order in $g$ we have:
    \begin{align}
      z_{m,J} & =  3g^2 \Gamma(d/4)^4
        \frac{\Gamma(J+m) \Gamma( m+1 - \frac{d}{2} ) }
        { \Gamma(\frac{d}{2}+J+m)  \Gamma(m+1 ) \Gamma(-\frac{d}{2}) \Gamma(\frac{d}{2} +1 )} + 
     {\cal O}(g^4) \crcr
         h_{m,J}& = \frac{d}{2} + J + 2m + 2 \frac{ 3g^2 \Gamma(d/4)^4 
     \Gamma(m + J)  \Gamma(m+1 - \frac{d}{2}) \sin\left( -\frac{\pi d}{2} \right)  }{ \Gamma(\frac{d}{2} + J + m ) \Gamma(m+1) \; \pi } + 
     {\cal O}(g^4) \;.
    \label{eq:hmJSol}
    \end{align}

 \paragraph{The OPE coefficients.} 
In appendix \ref{app:measure}, we give a detailed computation of the measure and residue factors which are needed for the OPE coefficients (\ref{eq:OPE coef}).
Here we simply present the final result. Putting all factors together the OPE coefficients are:
\begin{equation}\begin{split}
 c_{0,0}^2 & = 
   2 
-   \, 4\,  z_{0,0}\bigg[   \Psi(d/2) +  \Psi(1) - 2\Psi(d/4)  \bigg]  
+ O(z_{0,0}^2) \crcr
& =   2 \pm  \sqrt{ -3g^2 }  1 \frac{4\, \Gamma(d/4)^2   }{\Gamma(\frac{d}{2})} \bigg[ 2\Psi(d/4)  -  \Psi(d/2) -  \Psi(1)  \bigg]  
+ {\cal O}(g^3) \;,
\end{split}
\end{equation}
and for $(m,J) \neq (0,0)$:
\begin{align}
 c_{m,J}^2 & = 2\,
  \frac{   \Gamma(J+\frac{d}{2})   \Gamma(J+m )
    \Gamma(\frac{d}{2} + J + 2m  -1)
    \Gamma( 1 + m - \frac{d}{2}  )\Gamma(\frac{d}{4} + J + m  )^2
 }{  \Gamma(J+1) \Gamma(m+1 )
  \Gamma( 1 + 2m - \frac{d}{2}  )\Gamma( \frac{d}{2} + J + m    ) 
  \Gamma(J + 2m  )\Gamma(\frac{d}{2} + 2J + 2m   -1)
  \Gamma( \frac{d}{4} -m  )^2
 } \nonumber \\
 &  \quad + {\cal O}(g^2) \;.
\end{align}

\paragraph{Summary of results.} The conclusions of the computation of the OPE coefficients at small $g$ are:
\begin{itemize}
 \item[-] at $g=0$ we recover the classical dimensions $h^{\rm classical}=d/2 +J+2m$.
 \item[-] at $g\neq 0$ we get the dimensions $h_{m,J}= d/2 + J +2m +2z_{m,J}$ with $z_{0,0}\sim \sqrt{-g^2}$ and $z_{m,J} \sim g^2$ for $(m,J)\neq (0,0)$. For $(m,J)\neq (0,0)$, $z_{m,J}$ is always real. $z_{0,0}$
 is real for purely imaginary coupling and purely imaginary for real coupling. This is true at all orders in $g$.
 \item[-] at order $g^0$ all the OPE coefficients $c_{m,J}$ are real,  $c_{m,J}^2>0$. This is reassuring as it means that  the free theory is unitary, which it is (from OS positivity).
 \item[-] the OPE coefficients $c_{m,J}$ with $(m,J)\neq (0,0)$ are always real, $c_{m,J}^2>0$, at all orders in $g$  because they are series with real coefficients in $z_{m,J}$ which in turn is a series with real coefficients in $g^2$.
 \item[-] the OPE coefficient $c_{0,0}$ is: 
    \begin{itemize}
     \item complex ($c_{0,0}^2$ has a non zero imaginary part) at all orders in $g$ for $g$ real,
     \item real ($c_{0,0}^2>0$) at all orders in $g$ for $g$ purely imaginary,
    \end{itemize}
   this is because $c_{0,0}$ is a series with real coefficients in $z_{0,0}$.
\end{itemize}

\subsection{The $d=3$ case}
\label{sec:d=3}
In this subsection, we focus on the $d=3$ case.
Setting $\Delta_{\phi}=d/4$ with $d=3$, we obtain the eigenvalues:
	\begin{equation}
		k(h,J) \, =  \, 3g^2 \, \Gamma(3/4)^4 \, \frac{\Gamma(\frac{3}{4}+\frac{J}{2}-\frac{h}{2})\Gamma(\frac{h}{2}+\frac{J}{2}-\frac{3}{4})}
		{\Gamma(\frac{9}{4}+\frac{J}{2}-\frac{h}{2})\Gamma(\frac{h}{2}+\frac{J}{2}+\frac{3}{4})} \, ,
	\end{equation}
and the measure:
	\begin{align}
		\mu_{3/4}^3(h,J) \, 
		&= \, \left( \frac{1+(-1)^J}{2} \right) \frac{ \Gamma(h-1)\Gamma(J+\frac{3}{2})\Gamma(3-h+J)\Gamma(\frac{h+J}{2})^2}
		{\Gamma(h-\frac{3}{2})\Gamma(J+1)\Gamma(\frac{3-h+J}{2})^2\Gamma(h+J-1)} \nonumber\\
		&\hspace{80pt} \times \frac{\Gamma(\frac{h}{2}+\frac{J}{2}-\frac{3}{4})\Gamma(\frac{3}{4}-\frac{h}{2}+\frac{J}{2})}
		{\Gamma(\frac{3}{4}+\frac{h}{2}+\frac{J}{2})\Gamma(\frac{9}{4}-\frac{h}{2}+\frac{J}{2})} \, .
	\end{align}

\begin{figure}[htb]
	\begin{center}
		\scalebox{0.55}{\includegraphics{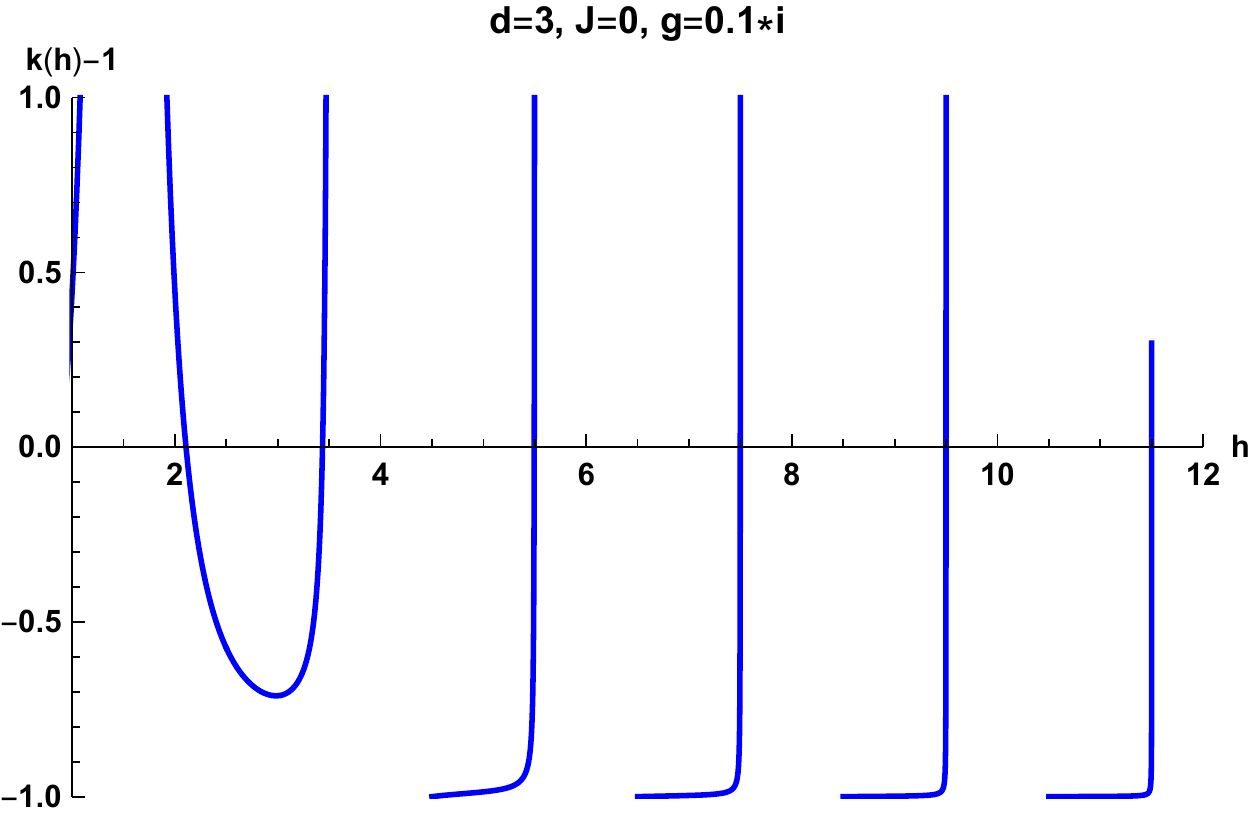}} \qquad 
		\scalebox{0.55}{\includegraphics{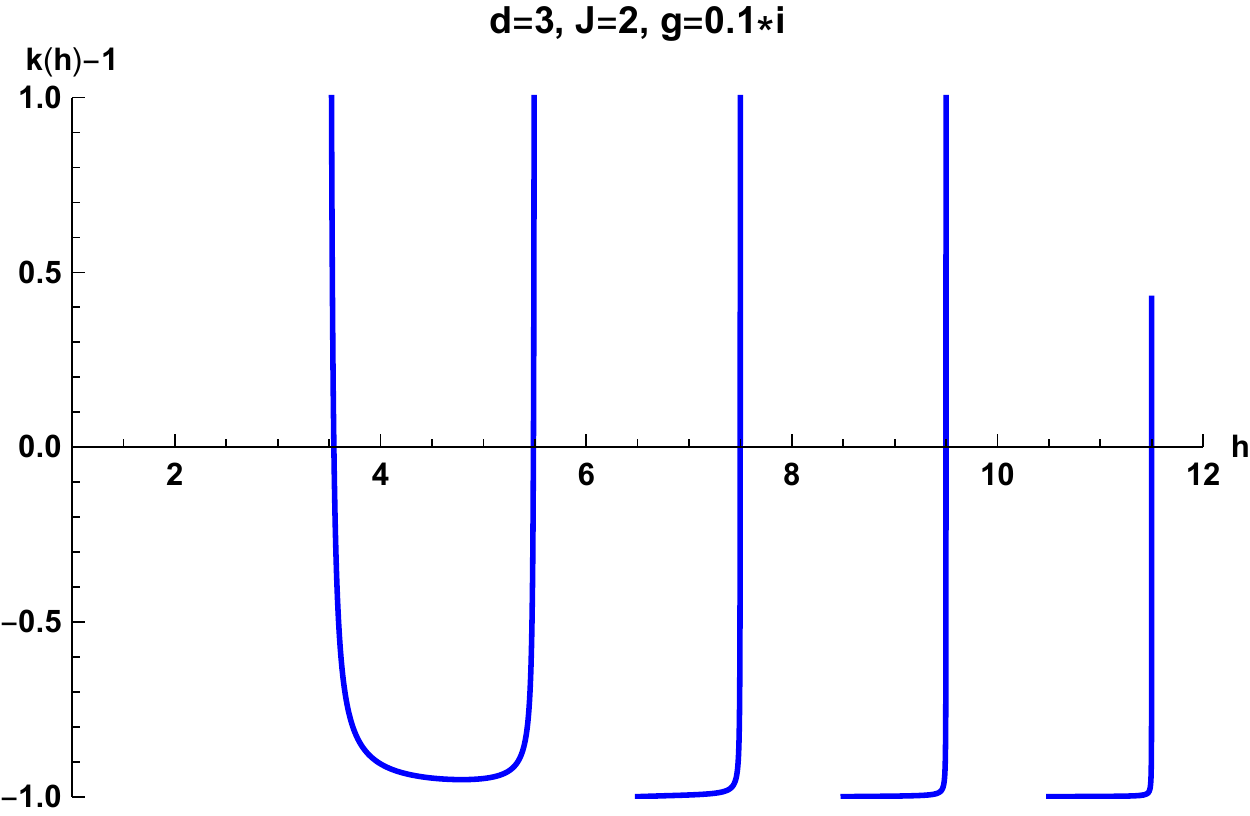}} \\ \, \vspace{30pt}
		\scalebox{0.55}{\includegraphics{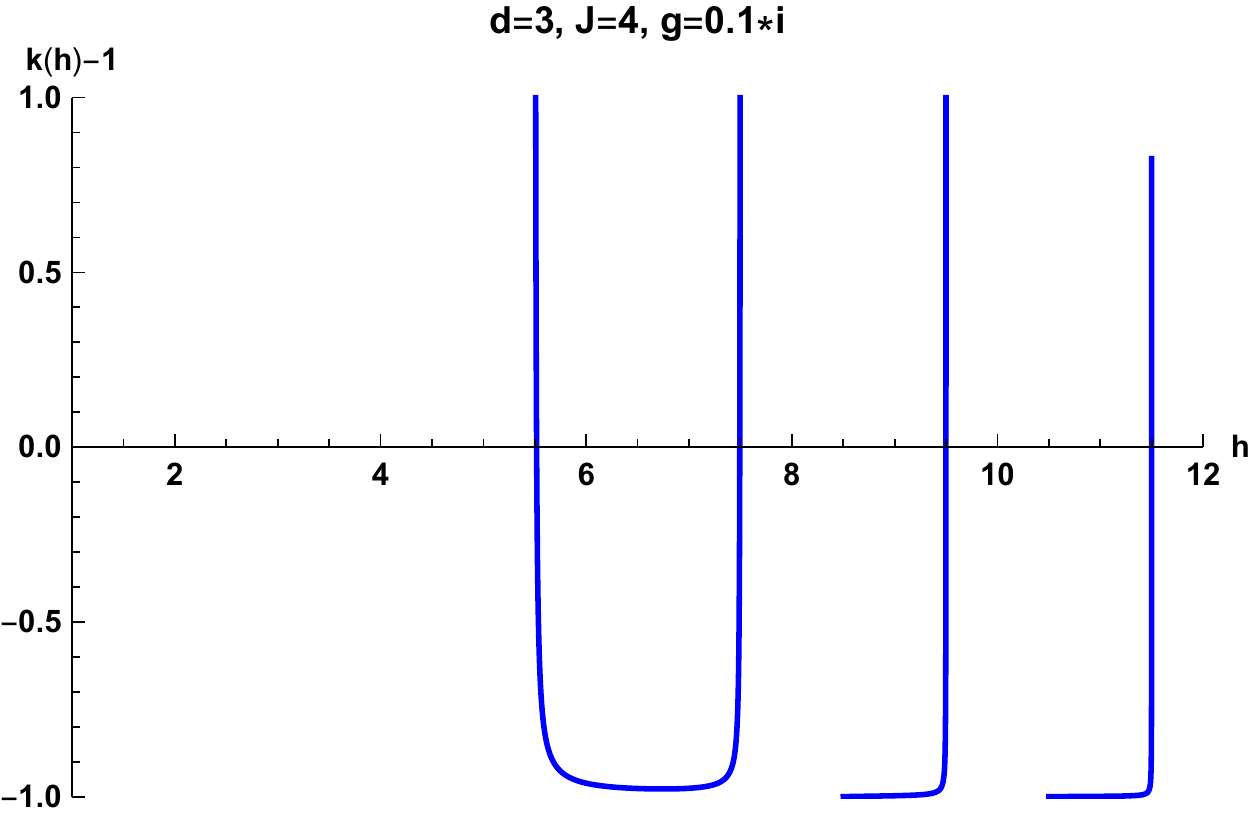}}
	\end{center}
	\caption{Plots of $k(h)-1$ for $d=3$, $g=0.1i$, and $J=0,2,4$.}
	\label{fig:d=3}
\end{figure}

The plots in Figure \ref{fig:d=3} graphically show that we can find one solution for the conformal dimension close to $h_{m,J}=3/2+J+2m$
for each non-negative integers $m$ and $J$.
For the $(m,J)=(0,0)$ case, there is a rather bigger deviation from $h_{0,0}=3/2$.

\paragraph{\it The case $(m,J)=(0,0)$.}
Expanding for small coupling constant $g$, we find the physical conformal dimensions:
	\begin{equation}
		h_{\pm} \, = \, \frac{3}{2} \, \pm \, 4 \sqrt{- \frac{3g^2}{\pi}} \, \Gamma(3/4)^2 \, + \, \mathcal{O}(g^3) \, ,
	\end{equation}
and associated OPE coefficients:
	\begin{equation}
		c_{\pm}^2 \, = \, 2 \, \pm \,   \frac{8}{\pi}(\pi - 2 - 4 \log 2)\Gamma(3/4)^2 \sqrt{-3\pi g^2} \, + \, \mathcal{O}(g^3) \, .
	\end{equation}
For a real value of the coupling constant $g$, both solutions $h_{\pm}$ are on the contour integral ${\rm Re}(h) = 3/2$  and  extra care is needed. In this case both $c_{\pm}^2$ are not real.

For a purely imaginary value of the coupling constant, $h_+$ is at the right of the contour while $h_-$ is at the left.
Therefore in this case only $h_+$ is in the spectrum of the model.
The associated OPE coefficient is $c_{+}^2=2 + \cdots>0$ for small coupling $g$.

\paragraph{\it The case $(m,J)\ne(0,0)$.}
The other solutions are also obtained by small coupling expansion as: 
	\begin{equation}
		h_{m,J} \, = \, \frac{3}{2} + J + 2m \, + \, \frac{6 \Gamma(\tfrac{3}{4})^4 \Gamma(m-\frac{1}{2}) \Gamma(m+J) }{\pi \, \Gamma(m+1)\Gamma(m+J+\frac{3}{2})}g^2
		\, + \, \mathcal{O}(g^4) \, , 
	\end{equation}
for non-negative integers $m$ and $J$ excluding the $(m,J)=(0,0)$ case.
The associated OPE coefficients are given by:
	\begin{equation}
		c_{m,J}^2 \, = \frac{ \Gamma(m-\frac{1}{2})\Gamma(m+\frac{1}{4})\Gamma(J+\frac{3}{2})\Gamma(m+J)\Gamma(m+J+\frac{3}{4})\Gamma(2m+J+\frac{1}{2})}
		{2^{4m+2J-2}\, \pi\,\Gamma(m+1)\Gamma(J+1)\Gamma(m-\frac{1}{4})\Gamma(2m+J)\Gamma(m+J+\frac{1}{4})\Gamma(m+J+\frac{3}{2})}
		\, + \, \mathcal{O}(g^2) \, .
	\end{equation}
The zeroth order (i.e. $g^0$) contributions are real and positive for any $m$ and $J$.
Hence, the OPE coefficients are real for all $m$ and $J$ in the small coupling regime.
This is a strong indication of unitarity of the model for $d=3$.

\subsection{The $d=2$ case}
\label{sec:d=2}
We  now focus on the $d=2$ case. We first set $d=2$ directly in the above equations. With $\Delta_{\phi}=d/4$, we have for the eigenvalues of the four point kernel: 
	\begin{equation}
		k(h,J) \, = \, - \, \frac{12 \pi^2 g^2}{(h+J-1)(h-J-1)} \, ,
	\end{equation}
therefore the solutions of $k(h,J)=1$ are given by:
	\begin{equation}
		h_{\pm} \, = \, 1 \, \pm \, \sqrt{J^2-12 \pi^2 g^2} \, .
	\end{equation}
A physical dimension needs to have real part at least $1$, therefore these solutions exist in the spectrum only in the following range of the coupling constant:
	\begin{align}
		&{\rm For\ \, } h_+: \quad g^2 \, \le \, \frac{J^2}{12 \pi^2} \, , \\
		&{\rm For\ \, } h_-: \quad \frac{J^2}{12 \pi^2} \, \le \, g^2 \, .
	\end{align}
From this we can see that $J=0$ is a special case where the $g^2 \to 0$ limit is well-defined for both $h_{\pm}$.
On the other hand, for $J>0$, the weak coupling limit $g^2 \to 0$ is only well-defined for $h_+$ and in the weak coupling limit $h_-$ does not exist in the spectrum.
Therefore, the $h_+$ solution in $J>0$ corresponds to the $h_{m,J}$ solution in Eq. ~\eqref{eq:hmJSol} with $m=0$.
The measure in $d=2$ is given by:
	\begin{equation}
		\mu_{1/2}^2(h,J) \, 
		= \, \left( \frac{1+(-1)^J}{2^{2h-2}} \right) \frac{\Gamma(\frac{h+J}{2})\Gamma(\frac{1-h+J}{2})}{\Gamma(\frac{h+J+1}{2})\Gamma(\frac{2-h+J}{2})} \, .
	\end{equation}

\paragraph{\it Spin $J=0$.}
The OPE coefficients for the $(m,J)=(0,0)$ case are:
	\begin{equation}
		c_{\pm}^2 \, = \,  2 \, \mp \sqrt{-3g^2}  \, 16 \pi \log2 \, + \, \mathcal{O}(g^2) \, ,
	\end{equation}
that is these OPE coefficients are real for small pure imaginary coupling $g$.

\paragraph{\it Spin $J>0$.}
The OPE coefficients for the $J>0$ case are: 
	\begin{equation}
		c_{0,J}^2 \, = \, \frac{2^{ 2-2J} \, \Gamma(J+\frac{1}{2})}{\sqrt{\pi}\, \Gamma(J+1)} \, + \, \mathcal{O}(g^2) \, .
	\end{equation}
The zeroth order $\mathcal{O}(g^0)$ contributions are real and positive for any $J$.
Hence, the OPE coefficients are real for all $J$ in the small coupling regime.

\subsection{Discontinuity at $d=2$}
\label{sec:discontinuity at d=2}
\begin{figure}[htb]
	\begin{center}
		\scalebox{0.65}{\includegraphics{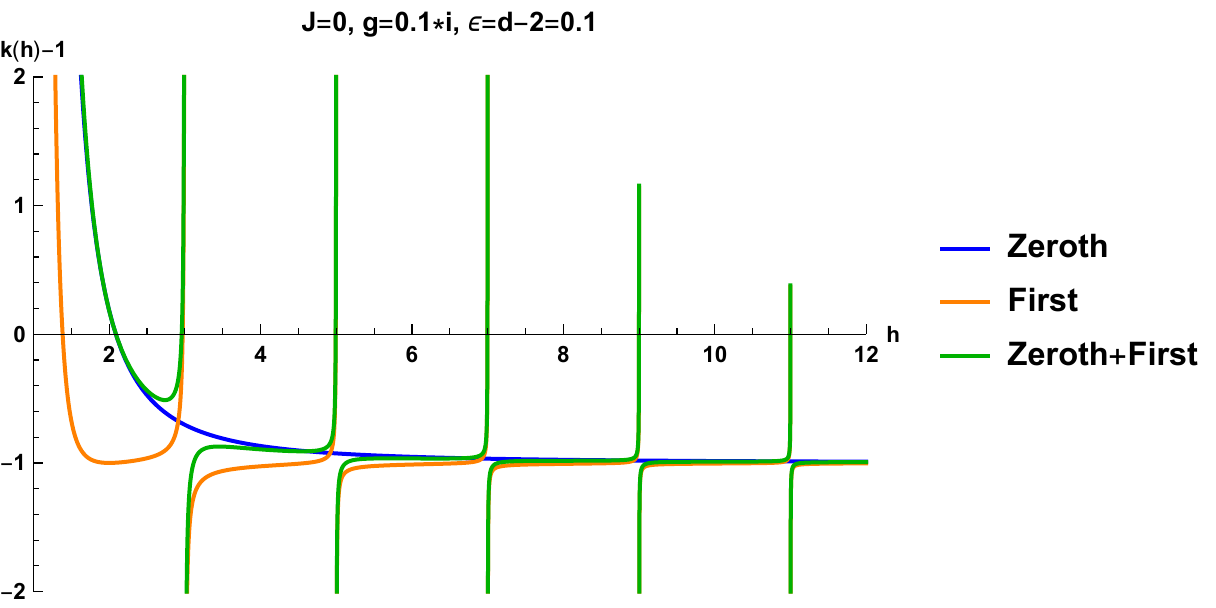}}  \qquad 
		\scalebox{0.65}{\includegraphics{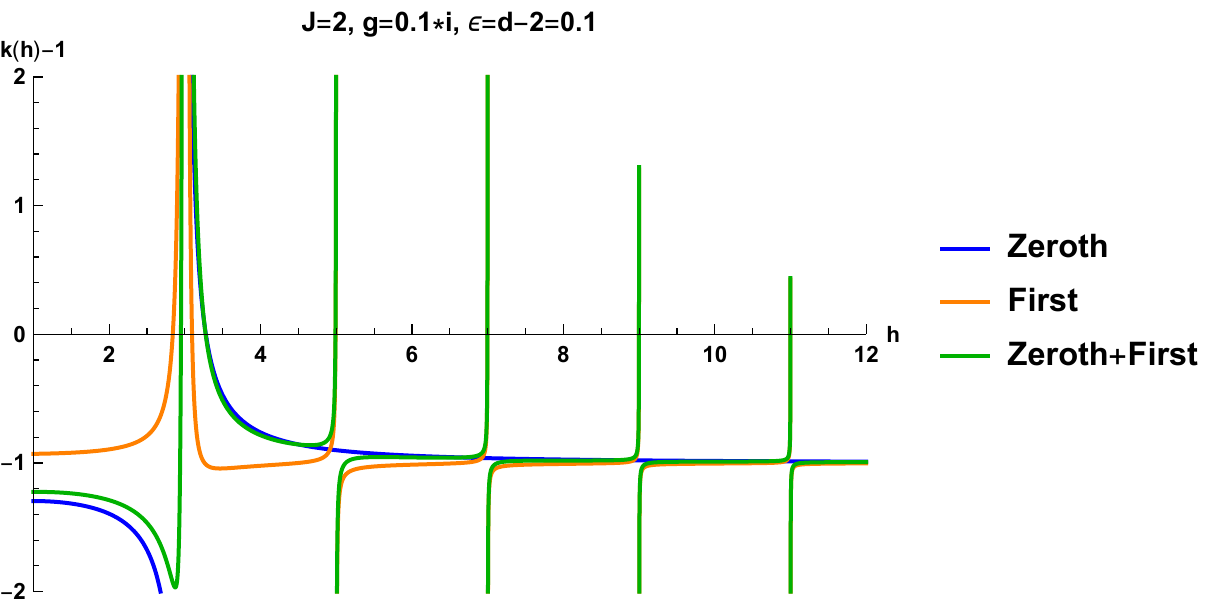}}  \\ \, \vspace{30pt}
		\scalebox{0.65}{\includegraphics{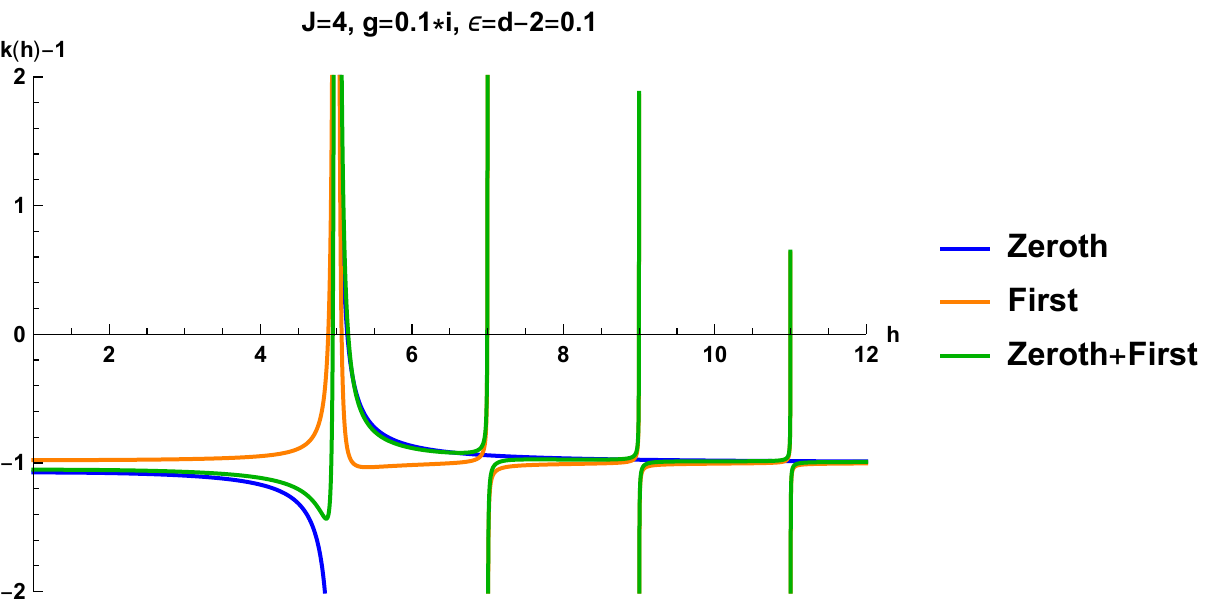}}
	\end{center}
	\caption{Plots of $k(h)-1$ around $d=2$ with $g=0.1i$, and $J=0,2,4$.
	The blue lines represent the result strictly at $d=2$ and the orange lines represent the first order correction of $\epsilon \equiv d-2$.
	For the plots, we took $\epsilon = 0.1$. The green lines are the results with the zeroth and first order corrections together.}
	\label{fig:d=2}
\end{figure}

At $d=2$, all solutions with $m>0$ in Eq.~\eqref{eq:hmJSol} disappear from the spectrum.
In order to better understand this phenomenon, let us consider $d=2+\varepsilon$ and expand the eigenvalue \eqref{eq:k(h,J)} in $\varepsilon$.
This leads to 
	\begin{align}
		k(h,J) \, = \, - \, \frac{12 \pi^2 g^2}{(h+J-1)(h-J-1)} \bigg[ 1 \, - \, \frac{1}{2} &\bigg( 2\gamma + \frac{1}{h+J-1} + \frac{3}{1-h+J} + 4\log 2 \nonumber\\
		&\quad + \Psi\left(\frac{1-h+J}{2}\right) + \Psi\left(\frac{h+J-1}{2}\right) \bigg) \, \varepsilon \, + \, \mathcal{O}(\varepsilon^2) \bigg] \, ,
	\end{align}
where $\gamma$ is the Euler-Mascheroni constant.
The zeroth order in $\varepsilon$ is a monotonically decreasing function of $h$ for any $J$,
while the first digamma function appearing in the $\mathcal{O}(\varepsilon)$ order introduces an infinite number of divergences at
	\begin{equation}
		h \, = \, 1 + J + 2m \, \qquad (m=0,1,2 \cdots)
	\label{d=2+epsilon solution}
	\end{equation}
and this leads to the solutions (\ref{eq:hmJSol}) with $m>0$.
Figure \ref{fig:d=2} shows this behavior of the eigenvalue.
Taking $d=2+\epsilon$ in Eq.~\eqref{OPE-coef-generalmJ} 
and sending $\epsilon\to 0$ we conclude that 
all the OPE coefficients $c_{m,J}$ (including those with $m>0$) have finite, non zero limits when sending $d$ to $2$.  

In fact we will argue in section \ref{sec:character} that the correct spectrum of the free theory at $d=2$ is actually given by Eq.~\eqref{d=2+epsilon solution}.
This is based on the character decomposition of the free theory partition function which we will present in detail.

\subsection{The $d=1$ case}
\label{sec:d=1}

The measure computed in appendix \ref{app:measure} is not defined for $d=1$. The correct measure for $d=1$ is given by \cite{Maldacena:2016hyu}:
	\begin{equation}
		\mu(h) \, =  \, \frac{2h-1}{ \pi \tan(\pi h/2)} \frac{\Gamma(h)^2}{\Gamma(2h)} \frac{\alpha_0 k(h,0)}{2} \, ,
	\label{eq:d=1 measure}
	\end{equation}
with $\alpha_0=\frac{\pi}{3 \Gamma(1/4)^4g^2}$, and $k(h,0)$ given by Eq.~\eqref{eq:k(h,J)} at $d=1$.
The on-shell value of the conformal dimensions $h_m$ are, from section \ref{sec:OPE}:
\begin{align}
h_{\pm}&=\frac{1}{2}\pm 2 \sqrt{-\frac{3g^2}{\pi}}\,\Gamma(1/4)^2 +\mathcal{O}(g^3) \;,\\
h_m&=\frac{1}{2}+2m-\frac{6\Gamma(1/4)^4g^2}{m\pi} +\mathcal{O}(g^4) \;, \;\;\; m>0 \;.
\end{align}
In $d=1$, a physical dimension needs to be greater than $1/2$. Therefore, in the weak purely imaginary coupling limit $h_{-}$ does not exist in the spectrum. For weak real coupling, $h_{\pm}$ are on the contour ${\rm Re}(h) =1/2$.
The OPE coefficients are given by:
	\begin{align}
		c_{\pm}^2 \, &= \, 2 \, \mp \, (\pi + 4\log 2) \Gamma(1/4)^2 \,  4 \sqrt{-\frac{3g^2}{\pi}} \, + \, \mathcal{O}(g^3) \, , \\
		c_m^2 \, &= \, \frac{4}{\pi} \frac{\Gamma(2m+1/2)^2}{ \Gamma(4m+1)} \, + \, \mathcal{O}(g^2) \, .
	\end{align}
Therefore, the OPE coefficients are real for the free theory ($g=0$) and for small pure imaginary coupling.

As a final comment on the $d=1$ case, if we expand the general formulae \eqref{eq:H(h)} around $d=1$, we obtain:
	\begin{align}
		\mu_{d/4}^d(h,0) \, &= \, \frac{\sqrt{\pi} \, \tan(\frac{\pi}{4}(2h+1)) \Gamma(h)}{2^{2h-2} \, \tan(\pi h/2) \, \Gamma(\frac{1}{2}+h)} \, + \, \mathcal{O}(d-1) \, .
	\end{align}
This result does not agree with the expression given in 
Eq.~\eqref{eq:d=1 measure}.
The reason for this is that the complete set of $h$ in $d=1$ is not just the principal series ($h=1/2+i r$), but also contains the discrete modes ($h=2n$) \cite{Kitaev:2017hnr}.
Therefore Eq.~\eqref{eq:4pt} (expressing the four point function in terms of conformal partial waves) must me modified by adding contours around the discrete modes $h=2n$ and changing  the measure $\mu(h)$ accordingly.

\section{A group-theoretic derivation of the spectrum of bilinear operators in the  free theory}
\label{sec:character}

In the limit of vanishing coupling, our model \eqref{eq:action} reduces effectively to a vector model with $O(N^3)$ symmetry. The two models are still distinguishable if one chooses to impose a singlet constraint based on one or the other group but, as we are here studying only bilinear operators, the two models should be indistinguishable in the free limit. Since our spectrum is continuous for $g\to 0$ (see appendix \ref{app:free} for the computation directly at $g=0$), this raises the question of why it is parametrized by $J$ and $m$ rather than by just $J$, as in the usual vector model (see for example \cite{Giombi:2016ejx} for a review).

Furthermore, we would like to understand the apparent discontinuity at $d=2$, which is present both at finite $g$ (previous sections) and at $g=0$ (appendix \ref{app:free}).

In this section, we reconstruct the spectrum of the free theory by a different method, as a way to cross-check our results, and in particular shed some light on the two questions above. We follow the set of ideas which have been developed in a number of papers, in connection to the Hagedorn transition in gauge theories \cite{Sundborg:1999ue,Aharony:2003sx} and the AdS/CFT duality between vector models and higher-spin theory \cite{Shenker:2011zf,Jevicki:2014mfa,Giombi:2014yra}. Similar methods have also been applied to tensor models in \cite{Beccaria:2017aqc,Bulycheva:2017ilt,Choudhury:2017tax}. Since we are here interested in the free theory, and its spectrum of bilinear operators, the $O(N)^3$ symmetry of our model will play no role, and we can actually treat its free limit as a $O(N^3)$ vector model. The main difference to the usual vector model, which we wish to highlight, is the effect of the non-canonical kinetic operator of our model on the spectrum of bilinear operators.

The spectrum of operators of a CFT on $\mathbb{R}^d$, or equivalently of CFT states on $\mathbb{R}\times S^{d-1}$, can be encoded in the grand canonical partition function with singlet constraint on $S^1 \times S^{d-1}$, where the $S^1$ is understood as Euclidean thermal circle with period $\beta$. First, we introduce the canonical or single-particle partition function:
\be \label{eq:singleZ}
Z(q,\mu) \, = \, \Tr[q^\Delta \mu^{j_3}] \, ,
\ee
where $q=e^{-\beta}$ and $\mu=e^{-\Omega}$, with $\Omega$ being the chemical potential conjugate to the eigenvalues of Cartan elements of $S^{d-1}$, denoted by $j_3$, and the trace is over all the states  built out of the elementary field $\phi$, which transforms in a real representation $R$ of the symmetry group $\cG$ (we will consider either $U(N)$ or $O(N)$). In this section, $\Delta$ denotes the conformal dimension, i.e.\ the eigenvalue of the dilation operator, which as usual plays the role of Hamiltonian in the radial quantization picture.

The grand canonical or multi-particle free energy, without singlet constraint, is related to the single-particle partition function by
\be
F \, = \, - \ln \cZ(q,\mu) \, = \, -\Tr[\ln (1-q^\Delta \mu^{j_3})^{-1}] \, = \, - \sum_{m=1}^{\infty} \frac{1}{m} Z(q^m, \mu^m) \,.
\ee
Following \cite{Sundborg:1999ue,Aharony:2003sx,Shenker:2011zf,Giombi:2014yra}, imposing the singlet constraint amounts to writing the multi-particle partition function with an integral over the symmetry group $\cG$:
\be \label{eq:Z-singlet}
\begin{split}
\cZ_\cG(q,\mu) &= \int_\cG [dU] \exp \left\{ \sum_i \sum_{m=1}^{\infty}  \frac{1}{m} q^{m \Delta_i} \mu^{m j_{3,i}} \chi^{\cG}_{R} (U^m) \right\} \\
&= \int_\cG [dU] \exp \left\{  \sum_{m=1}^{\infty}  \frac{1}{m} Z(q^m, \mu^m) \chi^{\cG}_{R} (U^m) \right\} \\
&\equiv \exp \left\{ \sum_{m=1}^{\infty}  \frac{1}{m} Z_\cG(q^m, \mu^m) \right\}\,,
\end{split}
\ee
where $[dU]$ is the normalized Haar measure, and $\chi^{\cG}_{R} (U)$ is the character of the group element $U\in\cG$ in the representation $R$.
The two cases which are relevant for us are \cite{Beccaria:2017aqc}:
\begin{align}
\chi^{U(N)}_{N\oplus \bar{N}} (U) & = \tr(U) + \tr(U^{-1})\,, 
\\
\chi^{O(N)}_{N} (U) & = \tr(U) \,. 
\end{align}

The integral over the group can be done explicitly, and the result can be expressed in terms of characters of the conformal group, which are in fact the single-particle partition function without singlet constraint Eq.~\eqref{eq:singleZ}.  
We denote the character of the $(\Delta_{\phi},J)$ representation of the $SO(2,d)$ conformal group by $\chi^{(d)}_{(\Delta_{\phi},J)}(q,\mu)$.
For the $U(N)$  gauge symmetry case one finds \cite{Shenker:2011zf}:
\be \label{eq:Z-UN}
Z^{(d)}_{U(N)}(q,\mu) \, = \, \left( \chi^{(d)}_{(\Delta_\phi,0)}(q,\mu) \right)^2 \, .
\ee
For the $O(N)$ gauge symmetry case, the $O(N)$ gauge singlet condition introduces an additional term in the partition function \cite{Jevicki:2014mfa, Giombi:2014yra} as:
\begin{align} \label{eq:Z-ON}
Z^{(d)}_{O(N)}(q,\mu) \, = \, \frac12 \left( \chi^{(d)}_{(\Delta_{\phi},0)}(q,\mu) \right)^2 \, + \, \frac12 \, \chi^{(d)}_{(\Delta_{\phi},0)}(q^2,\mu^2) \, .
\end{align}

The derivation above is very generic, based just on representation theory (the integral over the group is the way to count the number of singlets in a product of representations), and thus it applies also to our model with a non-canonical dimension $\Delta_\phi$ for the elementary field. The appearance of $\Delta_\phi\neq \frac{d-2}{2}$ is in fact the only difference between our $Z^{(d)}_{\cG}(q,\mu)$ and those found in the literature, and we are going to explore the consequences of this difference.

Before moving on, we should point out a subtle aspect of the above discussion.
The group integral enforcing the singlet constraint is usually introduced in the partition function by gauging the global symmetry on the compact manifold $S^1 \times S^{d-1}$, in the limit of vanishing gauge coupling, or equivalently by coupling the theory to a flat connection $A_\mu=U^{-1}\partial_\mu U$ and integrating over it. The connection can be gauged away, except for the constant $A_0$ component which has a non-trivial holonomy on $S^1$. For the usual vector model one can then show \cite{Shenker:2011zf,Giombi:2014yra} that the integral over $A_0$ reduces to the group integral in Eq.~\eqref{eq:Z-singlet}.
In our case, the non-integer power of the Laplacian renders such path integral derivation more perilous.
Gauging can actually be done in the standard way, simply replacing the derivatives with covariant derivatives, as best seen by expressing our kinetic operator in terms of the heat kernel by an inverse Laplace transform:
\be
\begin{split}
S_{\rm free}[\phi]   &=   \frac{\Gamma(1+\zeta)}{2} \int_{\gamma-\im \infty}^{\gamma+\im \infty} \frac{ds}{2\pi\im} s^{-1-\zeta} \int d^dx \,\sqrt{g} \;   \phi_{\mba}(x) e^{ - s \partial^2}\phi_{\mba}(x)\\
&=   \frac{\Gamma(1+\zeta)}{2} \int_{\gamma-\im \infty}^{\gamma+\im \infty} \frac{ds}{2\pi\im}  \sum_{n\geq 0} \frac{s^{-1-\zeta+n}}{n!} \int d^dx \,\sqrt{g} \;   \phi_{\mba}(x) (- \partial^2)^n \phi_{\mba}(x) \, .
\end{split}
\ee
The replacement $\partial_\mu \to \partial_\mu + A_\mu$ then leads to a gauge-invariant action.
Promoting our kinetic operator to a Weyl-covariant operator is instead more problematic, and we are not aware of any such generalization for non-integer powers of a Laplacian.\footnote{For integer powers, such generalization is known as the GJMS operators \cite{GJMS}.}
We thus take Eq.~\eqref{eq:singleZ} and Eq.~\eqref{eq:Z-singlet} as our starting point, putting aside a proper path integral derivation.

\subsection{$d=3$}
\label{app:d=3}
For $d=3$, the long representation ($\Delta>J+1$ for $J\ge 1$ and $\Delta>1/2$ for $J=0$) of the character for $SO(2,3)$ is given by:
	\begin{equation}
		\chi^{(3)}_{(\Delta,J)}(q,\mu) \, = \, \frac{q^{\Delta} \sum_{j=-J}^J \mu^j}{(1-q) (1-q \mu) (1-q \mu^{-1})} \, .
	\end{equation}
The short representations are obtained by eliminating corresponding null states:
	\begin{equation}
		\chi^{(3)}_{(\frac12,0)}(q,\mu) \, = \, \chi^{(3)}_{(\Delta,0)}(q,\mu)\Big|_{\Delta=\frac12} \, - \, \chi^{(3)}_{(\frac52,0)}(q,\mu) \, = \, \frac{q^{1/2} (1+q) }{(1-q \mu) (1-q \mu^{-1})} \, ,
	\end{equation}
for $J=0$ and:
	\begin{equation}
		\chi^{(3)}_{(J+1,J)}(q,\mu) \, = \, \chi^{(3)}_{(\Delta,J)}(q,\mu)\Big|_{\Delta=J+1} \, - \, \chi^{(3)}_{(J+2,J-1)}(q,\mu)
		\, = \, \frac{q^{J+1} \left[(q-\mu)\mu^J +(1-q\mu) \mu^{-J} \right]}{(1-\mu)(1-q) (1-q \mu) (1-q \mu^{-1})} \, ,
	\end{equation}
for $J\ge 1$.

Let us first consider the $U(N)$ gauge symmetry case.
For the canonical dimension of the fundamental scalar $\Delta_{\phi}=1/2$, we find that:
	\begin{equation}
		Z^{(d=3)}_{U(N)}(q,\mu) \, = \, \left( \chi^{(3)}_{(\frac12,0)}(q,\mu) \right)^2 \, = \, \chi^{(3)}_{(1,0)}(q,\mu) \, + \, \sum_{J=1}^{\infty} \, \chi^{(3)}_{(J+1,J)}(q,\mu) \, .
	\end{equation}
This result is the well-known Flato-Fronsdal decomposition \cite{Flato:1978qz}, which was also generalized to any dimension in \cite{Dolan:2005wy}.
Next, we consider the $\Delta_{\phi}=d/4=3/4$ case.
For this case, we have:
	\begin{equation}
		Z^{(d=3)}_{U(N)}(q,\mu) \, = \, \left( \chi^{(3)}_{(\frac34,0)}(q,\mu) \right)^2 \, = \, \sum_{J=0}^{\infty} \sum_{m=0}^{\infty} \, \chi^{(3)}_{(\frac32+J+2m,J)}(q,\mu) \, .
	\end{equation}

For the $O(N)$ gauge symmetry case, following the same computation as above we find:
	\begin{align}
		Z^{(d=3)}_{O(N)}(q,\mu) \, &= \, \frac12 \left( \chi^{(3)}_{(\frac12,0)}(q,\mu) \right)^2 \, + \, \frac12 \, \chi^{(3)}_{(\frac12,0)}(q^2,\mu^2) \nonumber\\
		&= \, \chi^{(3)}_{(1,0)}(q,\mu) \, + \, \sum_{J=1}^{\infty} \, \chi^{(3)}_{(2J+1,2J)}(q,\mu) \, ,
	\end{align}
for $\Delta_{\phi}=(d-2)/2 = 1/2$, while in the case 
$\Delta_{\phi}= d/4=3/4$ we get:
	\begin{align}
		Z^{(d=3)}_{O(N)}(q,\mu) \, &= \, \frac12 \left( \chi^{(3)}_{(\frac34,0)}(q,\mu) \right)^2 \, + \, \frac12 \, \chi^{(3)}_{(\frac34,0)}(q^2,\mu^2) \nonumber\\
		&= \, \sum_{J=0}^{\infty} \sum_{m=0}^{\infty} \, \chi^{(3)}_{(\frac32+2J+2m,2J)}(q,\mu) \, .
	\end{align}
This agrees with the results we found in section \ref{sec:OPE} and \ref{sec:d=3}.

\subsection{$d=2$}
\label{app:d=2}
For $d=2$, the long representation of the character for $SO(2,2)$ is given by \cite{Dolan:2005wy}:
	\begin{equation}
		\chi^{(2)}_{(\Delta,J)}(q,\mu) \, = \, \frac{q^{\Delta}\mu^{J}}{(1-q\mu)(1-q/\mu)} \, , \qquad \qquad (\Delta > J)
	\end{equation}
The short representations are again obtained by eliminating corresponding null states:
	\begin{equation}
		\chi^{(2)}_{(J,J)}(q,\mu) \, = \, \chi^{(2)}_{(\Delta,J)}(q,\mu)\Big|_{\Delta=J} \, - \, \chi^{(2)}_{(J+1,J-1)}(q,\mu)
		\, = \, \frac{q^J \mu^J}{1-q \mu} \, .
	\end{equation}

For the canonical dimension of the free scalar $\Delta_{\phi}=0$, we have:
	\begin{equation}
		Z^{(d=2)}_{U(N)}(q,\mu) \, = \, \left( \chi^{(2)}_{(0,0)}(q,\mu) \right)^2 \, = \, \sum_{J=0}^{\infty} \, \chi^{(2)}_{(J,J)}(q,\mu) \, ,
	\end{equation}
and:
	\begin{equation}
		Z^{(d=2)}_{O(N)}(q,\mu) \, = \, \frac12 \left( \chi^{(2)}_{(0,0)}(q,\mu) \right)^2 \, + \, \frac12 \, \chi^{(2)}_{(0,0)}(q^2,\mu^2)
		\, = \, \sum_{J=0}^{\infty} \, \chi^{(2)}_{(2J,2J)}(q,\mu) \, .
	\end{equation}

For $\Delta_\phi=d/4=1/2$ case, the decompositions are given by: 
	\begin{equation}
		Z^{(d=2)}_{U(N)}(q,\mu) \, = \, \left( \chi^{(2)}_{(\frac12,0)}(q,\mu) \right)^2 \, = \, \sum_{n, \bar{n}=0}^{\infty} \, \chi^{(2)}_{(1+n+\bar{n}, \, n-\bar{n})}(q,\mu) \, ,
	\end{equation}
and:
	\begin{equation}
		Z^{(d=2)}_{O(N)}(q,\mu) \, = \, \frac12 \left( \chi^{(2)}_{(\frac12,0)}(q,\mu) \right)^2 \, + \, \frac12 \, \chi^{(2)}_{(\frac12,0)}(q^2,\mu^2)
		\, = \,\sum_{\substack{n, \bar{n}=0\\ n+\bar{n}={\rm even}}}^{\infty} \, \chi^{(2)}_{(1+n+\bar{n}, \, n-\bar{n})}(q,\mu) \, ,
	\end{equation}
where for $O(N)$ case the summations over $n$ and $\bar{n}$ are taken only for the combinations such that $n+\bar{n}$ is an even integer.
This condition can be explicitly implemented by introducing an additional parameter $a=\{0,1\}$ and parametrizing $n=2n'+a$ and $\bar{n}=2\bar{n}'+a$.
Then the partition function for $O(N)$ can be written as:
	\begin{equation}
		Z^{(d=2)}_{O(N)}(q,\mu) \, = \, \frac12 \left( \chi^{(2)}_{(\frac12,0)}(q,\mu) \right)^2 \, + \, \frac12 \, \chi^{(2)}_{(\frac12,0)}(q^2,\mu^2)
		\, = \, \sum_{a=0,1} \sum_{n'=0}^{\infty} \sum_{\bar{n}'=0}^{\infty} \, \chi^{(2)}_{(1+2a+2n'+2\bar{n}', \, 2n'-2\bar{n}')}(q,\mu) \, .
	\end{equation}
We note that if we introduce the conformal weight $h$ ($\bar{h}$) of the holomorphic (anti-holomorphic) sector by:
	\begin{equation}
		\Delta \, = \, h \, + \, \bar{h} \, , \qquad J \, = \, h \, - \, \bar{h} \, ,  
	\end{equation}
then the spectrum in terms of $(h, \bar{h})$ is given by:
	\begin{equation}
		h \, = \, \frac{1+2n}{2} \, , \qquad \bar{h} \, = \, \frac{1+2\bar{n}}{2} \, .
	\end{equation}
	
For a free scalar field, as in this case, the symmetric spectrum between ($h, \bar{h}$) is expected.

The spectrum identifies here is not the one we found in section \ref{sec:d=2} by setting $d=2$. In fact, in order to reproduce the values $h+\bar h = 1 + J + 2\bar{n}$ we need to include not only the states with $m=0$ from section \ref{sec:d=2}, but also the states with $m>0$ which appear at $d=2+\epsilon$, Eq.~\eqref{d=2+epsilon solution}.

\subsection{$d=1$}
\label{app:d=1}
For $d=1$, there is no angular momentum or spin, so the character of the $SL(2,R)$ representation with weight $\Delta$ is given by:
	\begin{equation}
		\chi^{(1)}_{(\Delta)}(q) \, = \, \frac{q^{\Delta}}{1-q} \,.
	\end{equation}
Therefore, for $U(N)$ gauge symmetry case, we have
\footnote{The canonical dimension of the free boson in $d=1$ gives $\Delta = -1/2$ and unitary representation of $d=1$ does not exist for $\Delta < 0$ \cite{Dolan:2005wy}.
Nevertheless, if we naively use the above character formula for this canonical dimension, still the decomposition works as we show below.}
:
	\begin{equation}
		Z^{(d=1)}_{U(N)}(q) \, = \, \left( \chi^{(1)}_{(-1/2)}(q) \right)^2 \, = \, \sum_{m=0}^{\infty} \, \chi^{(1)}_{(m-1)}(q) \, ,
	\end{equation}
and:
	\begin{equation}
		Z^{(d=1)}_{U(N)}(q) \, = \, \left( \chi^{(1)}_{(1/4)}(q) \right)^2 \, = \, \sum_{m=0}^{\infty} \, \chi^{(1)}_{(m+\frac{1}{2})}(q) \, .
	\end{equation}
The former corresponds to a free scalar with the canonical dimension and the latter corresponds to the generalized free scalar with $\zeta=1/4$.

For the $O(N)$ gauge symmetry case, we obtain:
	\begin{equation}
		Z^{(d=1)}_{O(N)}(q) \, = \, \frac12 \left( \chi^{(1)}_{(-1/2)}(q) \right)^2 \, + \, \frac12 \, \chi^{(1)}_{(-1/2)}(q^2) \, = \, \sum_{m=0}^{\infty} \, \chi^{(1)}_{(2m-1)}(q) \, ,
	\end{equation}
and:
	\begin{equation}
		Z^{(d=1)}_{O(N)}(q) \, = \, \frac12 \left( \chi^{(1)}_{(1/4)}(q) \right)^2 \, + \, \frac12 \, \chi^{(1)}_{(1/4)}(q^2) \, = \, \sum_{m=0}^{\infty} \, \chi^{(1)}_{(2m+\frac{1}{2})}(q) \, .
	\end{equation}
This agrees with what we found in section \ref{sec:d=1}.

\section{Conclusion}
\label{sec:conclusions}

We studied the tensor model of \cite{Benedetti:2019eyl}, that is the  $O(N)^3$ with a modified free part:
\be
 S^{\rm free}[\phi]  =   \frac{1}{2} \int d^dx \;   \phi_{\mba}(x) (   - \partial^2)^{\zeta}\phi_{\mba}(x) \;, \qquad \zeta =d/4.
\ee
in $d<4$. The free theory is unitary. The conformal dimensions 
of the bilinear primary operators with arbitrary spin are given by $h_{m,J}=d/2 + J + 2m$ with $m\ge 0$.

Once we turn on a small tetrahedral coupling we obtain fixed points. One of them is infrared attractive for imaginary tetrahedral coupling. Near the fixed points, for $d\ne 2$, the conformal dimensions are shifted from the free value
by $\mathcal{O}(g^2)$ for $(m,J) \neq (0,0)$ and by 
$\mathcal{O}(\sqrt{-g^2})$ for $(m,J)=(0,0)$. The OPE coefficients are real and shifted by $\mathcal{O}(g^2)$ for $(m,J) \neq (0,0)$. 
For $(m,J)=(0,0)$, the OPE coefficient is shifted by $\mathcal{O}(\sqrt{-g^2})$. It stays real for imaginary tetrahedral coupling, but becomes complex for real tetrahedral coupling.

The model at $d=2$ is very special, and still unclear. While direct computation both in the free and interacting cases (sections \ref{sec:d=2} and appendix \ref{app:free}) seem to suggest that all the states with $m>0$ are absent at $d=2$, a derivation of the spectrum of the free theory based on character decomposition (section \ref{sec:character}) suggest that these states are in fact present. A deeper understanding of this point remains elusive. 
Inspired by the character decomposition it seems more natural to regard $d=2$ as the limit $\epsilon\to 0$ of $d=2+\epsilon$.  

We note that the spectrum of operators we have found does not include a spin-2 operator of dimension $d$. One could naively expect such an operator to exist, as it would correspond to the energy-momentum tensor of the theory.
The fundamental reason for its absence is that in our model the energy-momentum tensor, {\it if it exists}, is a non-local operator.

\bigskip

In the case of a purely imaginary tetrahedral coupling we have an infrared attractive fixed point. In this case (at all orders in the coupling) all the OPE coefficients of a bilinear primary operator and two fundamental fields are real.
Even though we have not exhausted {\it all} the primary operators in the model, our result is a strong indication that the 
large $N$ CFT at the infrared attractive fixed point 
is unitarity.

\section*{Acknowledgements}

We would like to thank Igor Klebanov for discussions on the conformal invariance in the long range Ising model and for pointing out references \cite{Paulos:2015jfa,Behan:2017emf}. We thank Nicolas Delporte for spotting a mistake in the normalization factor of the measure in the published version of the paper.

The work of DB, RG and SH is supported by the European Research Council (ERC) under the European Union's Horizon 2020 research and innovation program (grant agreement No818066).
The work of KS is supported by the European Research Council (ERC) under the European Union's Horizon 2020 research and innovation program (grant agreement No758759). 

This work was partly supported by Perimeter Institute for Theoretical Physics. Research at Perimeter Institute is supported by the Government of Canada 
through the Department of Innovation, Science and Economic Development Canada and by the Province of Ontario through the Ministry of Research, Innovation and Science.
During the completion of this work, DB has been hosted at
Laboratoire de Physique Th\'eorique (UMR 8627), CNRS,  Universit\'e Paris-Saclay, 91405 Orsay, France and SH was partly supported by \'{E}cole Normale Sup\'{e}rieure de Lyon, 46 All\'{e}e d'Italie, 69007 Lyon, France.

\appendix
\section{Measure and residue}
\label{app:measure}
In this appendix, we give a detailed computation of the measure and residues, which are needed for the computation of the OPE coefficients in section \ref{sec:OPE}.

 \paragraph{The measure.}
 We want to compute the measure at the physical dimensions $h_{m,J}= d/2 + J + 2m + 2z_{m,J}$. In this subsection, the results are valid for $d \neq 1,2$.
 From now on we consider only even spin, as otherwise the measure 
 in eq.~\eqref{eq:measure} is zero. Taking into account that  
 $\Delta_{\phi}=d/4$ the measure simplifies to:
\begin{equation}\label{eq:H(h)}
\begin{split}
 \mu_{d/4}^d(h, J)  & =    
   \frac{  \Gamma(J+\frac{d}{2}) 
  } {  \Gamma(J+1)} 
  \;  H^d_J(h) \; , \crcr
   H^d_J(h) & = \frac{
    \Gamma(\frac{ -\frac{d}{2}  +h+J}{2}) 
    \Gamma(\frac{ \frac{d}{2} -h+J}{2})
    \Gamma(h-1)\Gamma(d-h+J)\Gamma(\frac{h +J}{2})^2
 }{ 
  \Gamma(\frac{ \frac{3d}{2} -h+J}{2})\Gamma(\frac{\frac{d}{2} +h+J}{2}) 
  \Gamma(h-\frac{d}{2})\Gamma(h+J-1)\Gamma(\frac{d- h  +J}{2})^2
 }  \; .
 \end{split}
\end{equation}
We parametrize $h = d/2+J+2m + 2z$  and $H^d_J(d/2+J+2m + 2z) $ becomes:
\begin{equation} 
   \frac{ \Gamma(J+m+z)\Gamma( -m-z) 
    \Gamma(\frac{d}{2} + J + 2m +2z -1)
    \Gamma(\frac{d}{2} -2m -2z )\Gamma(\frac{d}{4} + J + m+z )^2
 }{ 
  \Gamma( \frac{d}{2} -m-z )\Gamma( \frac{d}{2} + J + m +z  ) 
  \Gamma(J + 2m +2z)\Gamma(\frac{d}{2} + 2J + 2m +2z -1)
  \Gamma( \frac{d}{4} -m-z )^2
 }  \; . 
\end{equation}

As the anomalous dimensions $z_{m,J}$ are small at small coupling, we can compute the measure at $h_{m,J}$ as Laurent series in $z_{m,J}$. 
Recalling that $\Gamma'(z) = \Gamma(z) \Psi(z) $ with $\Psi(z)$ the digamma function, we again have two cases.

\paragraph{\it The case $(m,J) = (0,0)$} The Laurent series of $ H^d_0 (d/2 +  2z) $ at small $z$ is obtained as:    
 \begin{equation}
    \begin{split}
 &  \frac{ \Gamma( z)\Gamma(  -z) 
    \Gamma(\frac{d}{2}   -2z )\Gamma(\frac{d}{4}  +z )^2
 }{ 
  \Gamma( \frac{d}{2}  -z )\Gamma( \frac{d}{2}  +z  ) 
  \Gamma( 2z)  \Gamma( \frac{d}{4} -z )^2 }  
  = \left( -\frac{2}{z}\right) \; 
  \frac{ \Gamma(1+ z) \Gamma(1 -z) 
    \Gamma(\frac{d}{2}   -2z )\Gamma(\frac{d}{4}  +z )^2
 }{ 
  \Gamma( \frac{d}{2}  -z )\Gamma( \frac{d}{2}  +z  ) 
  \Gamma(1+ 2z) 
  \Gamma( \frac{d}{4} -z )^2
 }  \crcr
 & \qquad \qquad  =\left(  -\frac{2}{z} \right) \bigg[\frac{1}{\Gamma(d/2)}   + z  \frac{1}{\Gamma(d/2)}  \bigg( 4\Psi(d/4) - 2 \Psi(d/2) - 2 \Psi(1)  \bigg) \bigg]
    +O(z) \;,
     \end{split}
    \end{equation}
therefore:
\begin{equation}
\begin{split}
 \mu_{d/4}^d \left( \frac{d}{2} +2 z ,  0 \right)  = -  \frac{2}{z}
+  4 \bigg[   \Psi(d/2) +  \Psi(1) - 2\Psi(d/4)  \bigg]  + O(z) \;. 
\end{split}
\end{equation}

\paragraph{\it The case $(m,J) \neq (0,0)$} Using $\Gamma(-m-z) \Gamma(1 + m + z) = (-1)^{m+1} \Gamma(1+z) 
 \Gamma(1-z) / z $ we have:
  \begin{equation} 
 \begin{split}
& H^d_J(d/2+J+2m + 2z) = \crcr
& \qquad  =     \frac{  (-1)^{m+1}    \Gamma(J+m )
    \Gamma(\frac{d}{2} + J + 2m  -1)
    \Gamma(\frac{d}{2} -2m   )\Gamma(\frac{d}{4} + J + m  )^2
 }{ z\Gamma(m+1 )
  \Gamma( \frac{d}{2} -m  )\Gamma( \frac{d}{2} + J + m    ) 
  \Gamma(J + 2m  )\Gamma(\frac{d}{2} + 2J + 2m   -1)
  \Gamma( \frac{d}{4} -m  )^2
 }  + O(z^0) \; .
\end{split}
 \end{equation}
In order to include as much as possible explicitly positive terms, it is convenient to use:
\begin{align*}
 (-1)^{m+1} \frac{ \Gamma(\frac{d}{2} -2m   ) }{   \Gamma( \frac{d}{2} -m  )  }  =  -\frac{\Gamma\left(1 + m-\frac{d}{2} \right) }{\Gamma\left(1 + 2 m-\frac{d}{2} \right)} \;,
\end{align*}
therefore:
\begin{align}
 & \mu_{d/4}^d \left( \frac{d}{2} + J + 2m + 2 z , J \right) =   
 (-1)  \frac{  \Gamma(J+\frac{d}{2}) 
  } { \Gamma(J+1)} 
 \\
 &  \quad \times
  \frac{    \Gamma(J+m )
    \Gamma(\frac{d}{2} + J + 2m  -1)
    \Gamma( 1 + m - \frac{d}{2}  )\Gamma(\frac{d}{4} + J + m  )^2
 }{ \Gamma(m+1 )
  \Gamma( 1 + 2m - \frac{d}{2}  )\Gamma( \frac{d}{2} + J + m    ) 
  \Gamma(J + 2m  )\Gamma(\frac{d}{2} + 2J + 2m   -1)
  \Gamma( \frac{d}{4} -m  )^2
 }  \;\frac{1}{z} + O(z^0) \; .  \nonumber
\end{align}

\paragraph{The $k'$ term.}
Next we need to evaluate $k'$ at $h_{m,J}$. Shifting to the $z$ variables, 
\begin{equation*}
k'(d/2 + J + 2m +2z,J ) = \frac{1}{2}\frac{d}{dz}k_{(m,J)}(z)\;,
\end{equation*}
where the functions $k_{(m,J)}(z)$ are defined  in eq.~\eqref{eq:2ks}.
We have:
\begin{align}
& \frac{d}{dz} k_{(0,0)}(z)  = k_{(0,0)}(z) \left[ -\frac{2}{z}  + \Psi(1+z) - \Psi(1-z)
 + \Psi\left(\frac{d}{2} -z \right)  -\Psi\left(\frac{d}{2}+z \right)\right] \\
& \frac{d}{dz} k_{(m,J)}(z)  \xlongequal{ (m,J) \neq (0,0) } k_{(m,J)} (z) \bigg[ -\frac{1}{z} + \Psi(J+m+z) 
  + \Psi(1+z) -\Psi(1-z) \crcr
  & \qquad + \Psi \left(m+1- \frac{d}{2} + z \right) 
-\Psi \left(\frac{d}{2} + J + m +z \right) -\Psi(m+1+z) 
  - \Psi\left( z-\frac{d}{2}\right) + \Psi\left( \frac{d}{2} +1 -z \right)
  \bigg] \;. \nonumber
\end{align}

At the physical dimension $z_{m,J}$, $k_{(m,J)}(z_{m,J})=1$ therefore we get the Laurent series:
\begin{equation}
 \frac{1}{2}\frac{d}{dz} k_{(0,0)}(z_{0,0}) 
    = -\frac{1}{z_{0,0}} + O(z_{0,0}) \;,\qquad
\frac{1}{2} \frac{d}{dz} k_{(m,J)}(z_{m,J}) = -\frac{1}{2 z_{m,J}} 
  + O(z_{m,J}^0)  \;.
\end{equation}
Observe that the Laurent series of $k_{(0,0)}$ does not have a constant term.

\section{The free theory for $\zeta\le1$}
\label{app:free}
The four point function in a (generalized) free CFT:
\be
  S[\phi]   =   \frac{1}{2} \int d^dx \;   \phi_{\mba}(x) (   - \partial^2)^{\zeta}\phi_{\mba}(x) \;,
\ee
with a real field of dimension $\Delta_{\phi} = d/2-\zeta$ can be written as in Eq.~\eqref{eq:4pt} with zero four point kernel:
\begin{equation}
 \langle{\phi(x_1) \phi(x_3) \rangle \; \langle \phi(x_2) \phi(x_4)} \rangle 
 +  \langle{\phi(x_1) \phi(x_4) \rangle \; \langle \phi(x_2) \phi(x_3)} \rangle 
   =  \sum_J 
  \int_{\frac{d}{2}-\imath \infty}^{\frac{d}{2}+\imath\infty} \frac{dh}{2\pi \imath}
   \; \mu_{\Delta_{\phi}}^d(h,J)
     G^{\Delta_{\phi}}_{h,J}(x_i)\;.
\end{equation}
The measure is given by Eq.~\eqref{eq:measure}:
\begin{align}\label{eq:measure1}
\mu_{\Delta_{\phi}}^d(h, J)  \, = & \, \left( \frac{ 1 + (-1)^J }{2} \right)
   \frac{  \Gamma(J+\frac{d}{2}) 
  }
  {  \Gamma(J+1)} 
   \crcr
&  \ \times    \frac{
 \Gamma( \frac{d}{2} - \Delta_{\phi})^2 
    \Gamma(\frac{ 2\Delta_{\phi} -d +h+J}{2}) 
    \Gamma(\frac{2\Delta_{\phi}-h+J}{2})
    \Gamma(h-1)\Gamma(d-h+J)\Gamma(\frac{h +J}{2})^2
 }{ \Gamma( \Delta_{\phi})^2 
  \Gamma(\frac{2d-2\Delta_{\phi}-h+J}{2})\Gamma(\frac{d-2\Delta_{\phi} +h+J}{2}) 
  \Gamma(h-\frac{d}{2})\Gamma(h+J-1)\Gamma(\frac{d- h  +J}{2})^2
 }  \;.
\end{align}

As in the case of the interacting theory, we can close the contour to the right and pick up the poles of the measure with ${\rm Re}(h)\geq d/2$, from which we should exclude the ``spurious'' poles of \cite{Simmons-Duffin:2017nub}, i.e.\ the poles of the measure that cancel with the poles of the conformal blocks.
Such spurious poles are the poles of the $\Gamma(d-h+J)$ factor in the numerator of Eq.~\eqref{eq:measure1}.
Since $h>d/2$, we are left with the poles of $\Gamma(\frac{2\Delta_{\phi}-h+J}{2})$, i.e.:
\be \label{eq:free-h}
h_{m,J} = 2 \Delta_{\phi} + J + 2 m \,,  \qquad m\in \mathbb{N} \;.
\ee
However, we have two Gamma functions in the denominator of 
Eq.~\eqref{eq:measure1} that can have poles, which, in the case that they coincide with any of the above poles, can lead to a zero residue (that is the absence of the corresponding pole).
The Gammas in  question are $\Gamma(\frac{2d-2\Delta_{\phi}-h+J}{2})$ and $\Gamma(\frac{d- h  +J}{2})^2$. Substituting Eq.~\eqref{eq:free-h} into Eq.~\eqref{eq:measure} we find:
\be
\text{Res}\left[\mu_{\Delta_{\phi}}^d(h,J) \right]_{h=h_{m,J} }  \propto \frac{1}{\Gamma(d-2\Delta_{\phi}-m) \Gamma(\frac{d}{2} -  \Delta_{\phi} -m)^2} \,.
\ee

For the canonical scaling, $\Delta_{\phi}=\frac{d}{2} -1$ (that is 
$\zeta=1$), the denominator is $\Gamma(2-m) \Gamma(1-m)^2$ and all the poles with $m\geq 1$ have zero residue. This means that the genuine poles are given by Eq.~\eqref{eq:free-h} with $m=0$. 
This spectrum coincides with the one of the vector model, not with the one of an interacting tensor model with standard propagator \cite{Giombi:2017dtl}.
This should not be a surprise, as the free theory is indistinguishable from a vector model with $O(N^3)$ symmetry, but as soon as interactions are turned on the symmetry is broken down to $O(N)^3$. 
Another way to understand this result is to notice that in the free theory any operator containing a factor $\partial^2\phi$ can be eliminated  by the equations of motion, regardless of the tensor rank.

For our scaling, $\Delta_{\phi}=\frac{d}{4}$ (that is $\zeta = d/4$), we find instead $\Gamma(\frac{d}{2}-m) \Gamma(\frac{d}{4}-m)^2$. Since the unitarity bounds require $d\leq 4$, we have three distinct cases: for $d=4$, we are back to the canonical case; for $d=2$, we have poles only from the first Gamma function, leading to the restriction $m=0$ as in Sec.~\ref{sec:d=2}; for $0<d<4$ and $d\neq 2$, neither of the two arguments of the Gamma functions are integers and all $m\geq 0$ are genuine poles.

While the discontinuity of the spectrum at $d=4$ can be understood as a consequence of the kinetic term becoming local, the discontinuity at $d=2$ remains puzzling. In fact, repeating the argument above, we would expect not to be able to remove operators with $\partial^2\phi$ factors for any $d<4$, as the Schwinger-Dyson equations:
\be
\langle \phi(x_1) \ldots \phi(x_n) (-\partial^2)^{d/4} \phi(x_0) \rangle = \sum_{i=1}^n \delta(x_0-x_i) \langle \prod_{j\neq i}^{1\ldots n} \phi(x_j)  \rangle \, ,
\ee
does not imply that we can eliminate $\partial^2 \phi$ inside correlation functions.
%

\section{The SYK model}
\label{app:original syk}
In this appendix, we review the OPE coefficients of the original SYK model and the conformal SYK model of Gross and Rosenhaus \cite{Gross:2017vhb}.

For the original SYK model, the OPE coefficients are given by \cite{Maldacena:2016hyu}:
	\begin{equation}
		c_m^2 \, = \, \alpha_0 \, \frac{(h_m-1/2)}{\pi \tan(\pi h_m / 2)} \frac{\Gamma(h_m)^2}{\Gamma(2h_m)} \, \frac{1}{k'(h_m)} \, , \qquad (m \, = \, 1, 2, \cdots) \;,
	\end{equation}
where:
	\begin{equation}
		\alpha_0 \, = \, 
		\frac{2\pi q}{(q-1)(q-2)\tan \frac{\pi}{q}} \, , 
	\end{equation}
and: 
	\begin{equation}
		k(h) \, = \, - \, (q-1) \, \frac{\Gamma(\frac{3}{2}-\frac{1}{q})\Gamma(1-\frac{1}{q})\Gamma(\frac{1}{q}+\frac{h}{2})\Gamma(\frac{1}{2}+\frac{1}{q}-\frac{h}{2})}
		{\Gamma(\frac{1}{2}+\frac{1}{q})\Gamma(\frac{1}{q})\Gamma(\frac{3}{2}-\frac{1}{q}-\frac{h}{2})\Gamma(1-\frac{1}{q}+\frac{h}{2})} \, .
	\end{equation}
The on-shell value of the conformal dimensions $h_m$ are determined by $k(h)=1$. Since $\tan(\pi h_m/2)<0$ and $k'(h_m)<0$, we have 
	\begin{equation}
		c_m^2 \, > \, 0 \, , \qquad (m \, = \, 1, 2, \cdots) \;.
	\end{equation}

The conformal SYK model considered by Gross and Rosenhaus \cite{Gross:2017vhb} has the OPE coefficients
	\begin{equation}
		c_m^2 \, = \, \alpha_0(q,\Delta) \, \frac{(h_m-1/2)}{\pi \tan(\pi h_m / 2)} \frac{\Gamma(h_m)^2}{\Gamma(2h_m)} \, \frac{1}{(1-2\bar{b})^2 k'(h_m)} \, , \qquad (m \, = \, 0, 1, 2, \cdots) \;,
	\end{equation}
with:
	\begin{equation}
		\alpha_0(q,\Delta) \, = \, \frac{2\pi}{(q-1)(1-2\Delta)\tan \pi\Delta} \, , 
	\end{equation}
and:
	\begin{equation}
		k(h) \, = \, - \, (q-1) \, \frac{\Gamma(\frac{3}{2}-\Delta)\Gamma(1-\Delta)\Gamma(\Delta+\frac{h}{2})\Gamma(\frac{1}{2}+\Delta-\frac{h}{2})}
		{\Gamma(\frac{1}{2}+\Delta)\Gamma(\Delta)\Gamma(\frac{3}{2}-\Delta-\frac{h}{2})\Gamma(1-\Delta+\frac{h}{2})} \, .
	\end{equation}
The on-shell value of the conformal dimensions $h_m$ are determined by $(1-2\bar{b})k(h)=1$.
The dependence of the coupling constant comes from $\bar{b}$ which is determined by
	\begin{equation}
		\frac{\bar{b}^q}{1-2\bar{b}} \, = \, \frac{1}{2\pi J^2} \, (1-2\Delta)\tan\pi \Delta \, .
	\label{eq:bbar}
	\end{equation}

For $\Delta=1/q$, the OPE coefficients are identical to those of the original SYK model except the $(1-2\bar{b})^{-2}$ factor.
For real value of the coupling constant, this factor is always positive. Therefore in this model, for any value of real coupling constant 
	\begin{equation}
		c_m^2 \, > \, 0 \, , \qquad (m \, = \, 0, 1, 2, \cdots)
	\end{equation}

Let us now explicitly compute the OPE coefficient for small coupling $|J|\ll 1$ in this model.
To compare with our model we set $q=4$ and $\Delta=1/q=1/4$.

First from Eq.(\ref{eq:bbar}), we can explicitly solve for $\bar{b}$ as
	\begin{equation}
		\bar{b} \, = \, \frac{1}{2} \, - \, \frac{\pi}{8} J^2 \, + \, \mathcal{O}(J^4) \, .
	\end{equation}
The solution of the conformal dimensions are now given by
	\begin{equation}
		h_m \, = \, \frac{3}{2} + 2m \, + \, \frac{3J^2}{4(1+2m)} \, + \, \mathcal{O}(J^4) \, ,
	\end{equation}
and 
	\begin{equation}
		c_m^2 \, = \, \alpha_0(4,1/4) \, \frac{3\Gamma(3/2+2m)^2}{\pi^2 \Gamma(3+4m)} \, + \, \mathcal{O}(J) \, .
	\end{equation}


\providecommand{\href}[2]{#2}\begingroup\raggedright\endgroup


\end{document}